%% file: symm_paper.tex
\documentclass[a4paper, 12pt]{article}
\usepackage[usenames]{color} 
\usepackage{amsmath,amssymb}
\numberwithin{equation}{section}
\usepackage[top=2.5cm,left=3cm,right=3cm,bottom=3cm,foot=1cm,]{geometry}
\usepackage[breaklinks, colorlinks, citecolor=blue]{hyperref}
\usepackage{multirow,array,graphicx}
\usepackage{sidecap,epstopdf,dcolumn}
\usepackage{bm}% bold math
\usepackage[font={small,it}]{caption}
\usepackage[footnotesize]{subfigure}
\usepackage[]{appendix}
\usepackage[numbers, square, comma,  sort&compress]{natbib}
\input{defs}

\begin{document}

\title{{\bf Simulating the symmetron:\\domain walls and   symmetry-restoring impurities}}
\author{\sc{Jonathan A. Pearson\footnote{E-mail: \href{mailto:jonathan.pearson@durham.ac.uk}{jonathan.pearson@durham.ac.uk}}}\\ \\ {\it Centre for Particle Theory}, \\{\it Department of Mathematical Sciences,} \\{\it Durham University, South Road, }\\{\it Durham, DH1 3LE, U.K.}}

\date{\today}

\maketitle
\begin{abstract}
In this paper we study the dynamics of relativistic domain walls in the presence of static symmetry-restoring impurities. The field theory is precisely the same as what is known to cosmologists as the ``symmetron model'', whereby the usual $\mathbb{Z}_2$ symmetry breaking potential is appended with a space-varying mass-term (the space-variation is set by the profile of the impurity, which we take to be a ``tanh''-function). After presenting the outcomes of a suite of different numerical experiments we have three main results: (1) domain walls pin to impurities, (2) domain wall necklaces can be energetically preferred configurations, and (3) impurities significantly modifies the usual $\qsubrm{N}{dw}\propto t^{-1}$ scaling law for random networks of domain walls.
\end{abstract}
\clearpage

\tableofcontents

\input{symmetron_content}

\addcontentsline{toc}{section}{Acknowledgements}

\section*{Acknowledgements}
The author appreciates comments from Clare Burrage, Peter Cuttell, Paul Saffin, Jeremy Sakstein, and Shuang-Yong Zhou, and  is supported by the STFC Consolidated Grant ST/J000426/1. The author would like to  thank the Lorentz Center at Leiden University for their  hospitality while this work was begun, and the kind hospitality of the folks at Keepers Cottage whilst the final draft of this article was prepared. 
\appendix
\input{symm_appendix}

\addcontentsline{toc}{section}{References}
\bibliographystyle{JHEP}
\footnotesize{
%\bibliography{refs}
\input{symm_paper.bbl}
}
\end{document}

%% file: defs.tex
\newcommand{\half}[0]{\frac{1}{2}}

\newcommand{\pd}[2]{\frac{\partial #1}{\partial #2}}
\newcommand{\ld}[0]{\mathcal{L}}

\newcommand{\dd}[0]{\textrm{d}}
\newcommand{\defn}[0]{\equiv}

\newcommand{\qsubrm}[2]{{#1}_{\scriptscriptstyle{\textrm{#2}}}}
\newcommand{\qsuprm}[2]{{#1}^{\scriptscriptstyle{\textrm{#2}}}}

\newcommand{\rbm}[1]{{\bf{#1}}}

\renewcommand{\figurename}{Figure}

\def\be{\begin{equation}}
\def\ee{\end{equation}}
\def\bea{\begin{eqnarray}}
\def\eea{\end{eqnarray}}
\def\bse{\begin{subequations}}
\def\ese{\end{subequations}}

\newcommand{\fref}[1]{{Figure \ref{#1}}}

\let\oldsqrt\sqrt
\def\sqrt{\mathpalette\DHLhksqrt}
\def\DHLhksqrt#1#2{%
\setbox0=\hbox{$#1\oldsqrt{#2\,}$}\dimen0=\ht0
\advance\dimen0-0.2\ht0
\setbox2=\hbox{\vrule height\ht0 depth -\dimen0}%
{\box0\lower0.4pt\box2}}

%% file: symmetron_content.tex
\section{Introduction}
The scaling dynamics of domain walls \cite{vs_top} have been extensively and systematically studied, using analytical \cite{Bray:1990zz, Bray:1994zz, Filipe:1994pc, Hindmarsh:1996xv} and numerical \cite{Oliveira:2004he, Battye:2006pf, Battye:2011ff, Leite:2011sc} techniques (this list of references is by no means complete, but hopefully represents a faithful cross-section through recent developments in the literature). The upshot is that the evolution of the number of domain walls scales as $\qsubrm{N}{dw}(t)\propto t^{-1}$. This scaling law can be thought of as describing the ``quickest possible'' collapse of a network of domain walls: when there is nothing to resist their collapse, domain walls shrink as fast as causality allows. There have been some proposals for the construction of a field theory with domain walls whose networks   violates the $\propto t^{-1}$ scaling law: domain walls which meet at junctions showed promise, but ultimately the networks still collapsed at the  $\propto t^{-1}$ rate  \cite{Battye:2006pf}. A proposal which does work, and has the desired effect of slowing down the wall evolution came after  the discovery of exact analytical solutions \cite{Battye:2008zh, Battye:2009nf} of  superconducting domain walls (known as ``kinky vortons''). Subsequent numerical experiments \cite{Battye:2009ac, Battye:2010dk} have shown that the presence of a conserved charge which is localised on the domain wall can impede the collapse of a wall network, thus violating the $\propto t^{-1}$ scaling law.

There are many studies in the field theory literature of the dynamics of domain wall networks in the simplest sense (see e.g., the above references as well as \cite{Thouless, Garagounis:2002kt, Pogosian:2002ua, Antunes:2003be, Oliveira:2004he, Avelino:2005pe, Avelino:2006xy, Avelino:2006xf, PinaAvelino:2006ia, Battye:2006pf, Avelino:2008ve, Avelino:2008qy, Avelino:2009tk, Sousa:2009is, Battye:2010dk, Battye:2011ff, Leite:2011sc, Leite:2012vn, Correia:2014kqa} for recent studies), but very few (if any) when the domain walls could interact with objects, and so it is interesting to understand what effect the presence of ``objects'' could have on the scaling dynamics of the wall network. There are studies of two-component Bose-Einstein condensates in the literature, e.g., \cite{Kasamatsu:2010aq, Kasamatsu:2013qia}, in which an interaction between domain walls and vortices is studied. We  acknowledge that domain walls and impurities have been studied in a non-relativistic context by \cite{PhysRevLett.54.2708}, but we have not been able to find any studies of relativistic domain walls in the presence of impurities. This apparent lack of understanding is the first motivation for the present work.

The second motivation is  a desire to further understand what is known to the cosmology community as the ``symmetron mechanism''. This is a model which was introduced   \cite{Pietroni:2005pv, Olive:2007aj, Hinterbichler:2010es, Khoury:2010xi} as a means of allowing a scalar degree of freedom to simultaneously satisfy local tests of General Relativity and mediate a long-range force  which contributes towards the acceleration of the  Universe. This symmetron model has been subject to some simulation in a cosmological context  \cite{Brax:2012nk, Llinares:2013jua, Llinares:2013qbh,  Gronke:2013mea, Silva:2013sla, Taddei:2013bsk}; there has also been a recent article studying the nonlinear dynamics of the Vainshtein mechanism \cite{Brito:2014ifa} (this is, in addition to the symmetron mechanism, an example of a ``screening mechanism''). The symmetron model   contains domain walls which disappear in  high density environments since there the spontaneously broken symmetry is restored.  During the final stages of preparation, we learnt of a similar paper being independantly prepared \cite{Levon_dw_symm}  whose results complement ours. The dynamics of ``pure domain walls'' are sufficiently complicated and non-trivial  in flat space that the symmetron model warrants systematic study as a field theory in its own right.

 In this paper we study the relativistic dynamics of domain walls  in the presence of static impurities. Specifically, we will present the results from a suite of numerical experiments of the  $\mathbb{Z}_2$ domain wall theory in $(2+1)$-dimensions where the scalar field $\phi$ has a quadratic interaction term with some fixed smooth density profile.

The layout of this paper is as follows. In the following subsections of the Introduction we describe the field theory, specify our functional form of the impurities, and give some numerically-obtained 1D solutions to the equations of motion. In Section \ref{sec:idealized} we construct some ``idealized'' configurations: these are very simple (again, numerically obtained) solutions to the equations of motion which will enrich the understanding of the model: the most interesting of which is the existence of \textit{domain wall necklaces} outlined in Section \ref{sec:dwneck}. In Section \ref{eq:sec;dyn-rand} we present results from ``random initial condition'' simulations: it is in this section where we present the scaling dynamics of domain wall networks in the presence of symmetry-restoring impurities\footnote{To supplement the results presented in this paper we have prepared a number of movies of the evolution which can be found at \url{http://www.jpoffline.com/symmetron}.}, and we find that $\qsubrm{N}{dw}\propto t^{-\gamma}$ with $\gamma<1$ in general when there are impurities. Final remarks are saved for Section \ref{sec:discussion}. In Appendix \ref{sec:append-numerics} we describe our numerical implementation.

\subsection{The model}
The theory we study is described by the canonical Lagrangian density for a real scalar field $\phi(t, \rbm{x})$,
\bse
\label{eq:symm-lag-pre-shuffle}
\bea
\ld = \half \partial_{\mu}\phi\partial^{\mu}\phi - V(\phi),
\eea
where the potential is given by
\bea
\label{eq:sec:symm_pot}
V(\phi) = \frac{\lambda}{4}\left( \phi^2 - \eta^2\right)^2 + \half \frac{\qsubrm{\rho}{m}}{M^2} \phi^2.
\eea
\ese
On the face of it there are three parameters in the theory, $\lambda, \eta$ and $M$. In the present work we will refer to $\qsubrm{\rho}{m} = \qsubrm{\rho}{m}(\rbm{x})$ as the \textit{impurity density}. 

In a field theoretic context, $\qsubrm{\rho}{m}$ corresponds to the profile of some massive field -- massive compared to the $\phi$-field; as an example, one could consider massive Q-balls. In a cosmological context, $\qsubrm{\rho}{m}$ corresponds to the density profile of matter, and the potential (\ref{eq:sec:symm_pot}) is the effective potential that the scalar $\phi$ feels in the Einstein frame. Although we will specify a particular functional form later on, we ask that the density of a single impurity  has the following asymptotic behavior for distances $r$ from its centre:
\bea
\lim_{r\rightarrow \infty}\qsubrm{\rho}{m} = 0,\qquad \lim_{r\rightarrow 0}\qsubrm{\rho}{m} = \rho_0.
\eea
When $\qsubrm{\rho}{m} < \lambda\eta^2M^2$ the $\mathbb{Z}_2$ symmetry is spontaneously broken in the vacuum and domain walls can be expected to exist.
When $\qsubrm{\rho}{m} >\lambda\eta^2M^2$ the $\mathbb{Z}_2$ symmetry is restored and there are no domain wall solutions. The mass scale $M$ is an important quantity for the cosmological application.   In this paper we are not too concerned about solving the field equations with a ``cosmologically'' viable hierarchy of scales  since we will be constrained by picking \textit{numerically viable} values -- this issue arises since large values of $M$ require prohibitively high resolution for the lattice methods we use to solve the field equations. That said, we expect   intuition to still be obtained.

It is simple to rescale the fields and spatial coordinates to provide a dimensionless theory.  By defining  
\bse
\bea
\tilde{x}^{\mu} \defn \eta\sqrt{\lambda} x^{\mu},\qquad \psi \defn \frac{\phi}{\eta},\qquad \hat{\rho}\qsubrm{ }{m}\defn \frac{\qsubrm{\rho}{m} }{M^2\lambda\eta^4}
\eea
the Lagrangian density (\ref{eq:symm-lag-pre-shuffle}) becomes
\bea
\frac{\ld}{\eta^4\lambda} = \half \tilde{\partial}_{\mu}\psi \tilde{\partial}^{\mu}\psi-  \frac{1}{4}\psi^4 + \half \left( 1-\hat{\rho}\qsubrm{ }{m} \right)\psi^2 - \frac{1}{4} .
\eea
\ese
We  use this freedom to set $\lambda = \eta=1$, meaning that the only scale left in the problem is the size of $\qsubrm{\hat{\rho}}{m}$. Notice that we are putting $M$ inside $\hat{\rho}\qsubrm{}{m}$.

The equation of motion derived from the theory (\ref{eq:symm-lag-pre-shuffle}) is
\bea
\label{eq:eom}
\ddot{\phi} - \nabla^2\phi + \frac{\dd V}{\dd\phi}=0,
\eea
and the energy is given by
\bea
\label{eq:sec:defn-energy}
E = \int \dd^dx\, \left[ \half \partial_i\phi\partial_i\phi + V(\phi)\right].
\eea
The integrand of (\ref{eq:sec:defn-energy}) is the energy density, which we denote as $\mathcal{E}(\rbm{x})$.
To solve the equation of motion we discretize onto a lattice, use finite differences to approximate derivatives, and use a leapfrog algorithm to update the values of the field. See Appendix \ref{sec:append-numerics} for a full explanation of our implementation.

For a substantial portion of the present work we will be interested in finding static solutions to the equations of motion: that is, obtaining the scalar field profile for a given impurity density that satisfy
\bea
\nabla^2\phi = \frac{\dd V}{\dd\phi}.
\eea
To do this we use a relaxation method known as \textit{gradient flow}.  The gradient flow technique works by  flowing ``gently'' towards the nearest minimum of the energy functional. There are more rigorous explanations of the technique available \cite{dur4361}, but it is a very successful strategy for constructing static solutions (see, e.g., \cite{Irwin:1996nj, Battye:2008mm, Battye:2013tka}). The scheme for obtaining such solutions can be obtained by ``evolving'' the equation
\bea
\dot{\phi} = \nabla^2\phi - \frac{\dd V}{\dd\phi},
\eea
and computing the value of the energy (\ref{eq:sec:defn-energy}). A static configuration is found when the change in energy $\dot{E}\approx 0$.   

\subsection{Specification of the impurity density}
\label{sec:impurity-prfiles}
We will take the impurity density to be a ``top-hat''  circularly symmetric profile whose precise form  is  specified by three parameters: the radius $r_0$,   skin thickness $s$, and internal density $\rho_0$. The functional form of a single impurity in polar coordinates is
\bea
\label{eq:sec:1profile}
\qsubrm{\rho}{m}(r) =\rho_0\half \left[ 1 - \tanh\left(\frac{r - r_0}{s}\right)\right].
\eea 
We will frequently study systems with multiple impurities; in Cartesian coordinates $(x,y)$ the total impurity density $\qsubrm{\rho}{m} = \qsubrm{\rho}{m}(x,y)$ corresponding to $N$ individuals is given by
\bea
\qsubrm{\rho}{m}{}(x,y) = \sum_{n=1}^N {\rho}_{n}(x,y; \mathcal{P}_n),
\eea
where $\mathcal{P}_n \defn \{{x}_{0,n}, {y}_{0,n},\rho_{0,n},r_{0,n},s_{n}\}$ are the set of parameters specifying the properties of the $\qsuprm{n}{th}$ impurity whose  density profile  is given by
\bea
{\rho}_{n}(x,y; \mathcal{P}_n) = \frac{\rho_{0,n}}{2}\left[ 1 - \tanh\left( \frac{\sqrt{(x-x_{0,n})^2 + (y-y_{0,n})^2}- r_{0,n}}{s_n} \right) \right].
\eea
Note that $({x}_{0,n}, {y}_{0,n})$ specifies the coordinates of the centre of the $\qsuprm{n}{th}$ impurity.
\subsection{1D static solutions}
\begin{figure}[!t]
      \begin{center}
      	 \subfigure[$\, \phi(x)$]{{\includegraphics[scale=0.6]{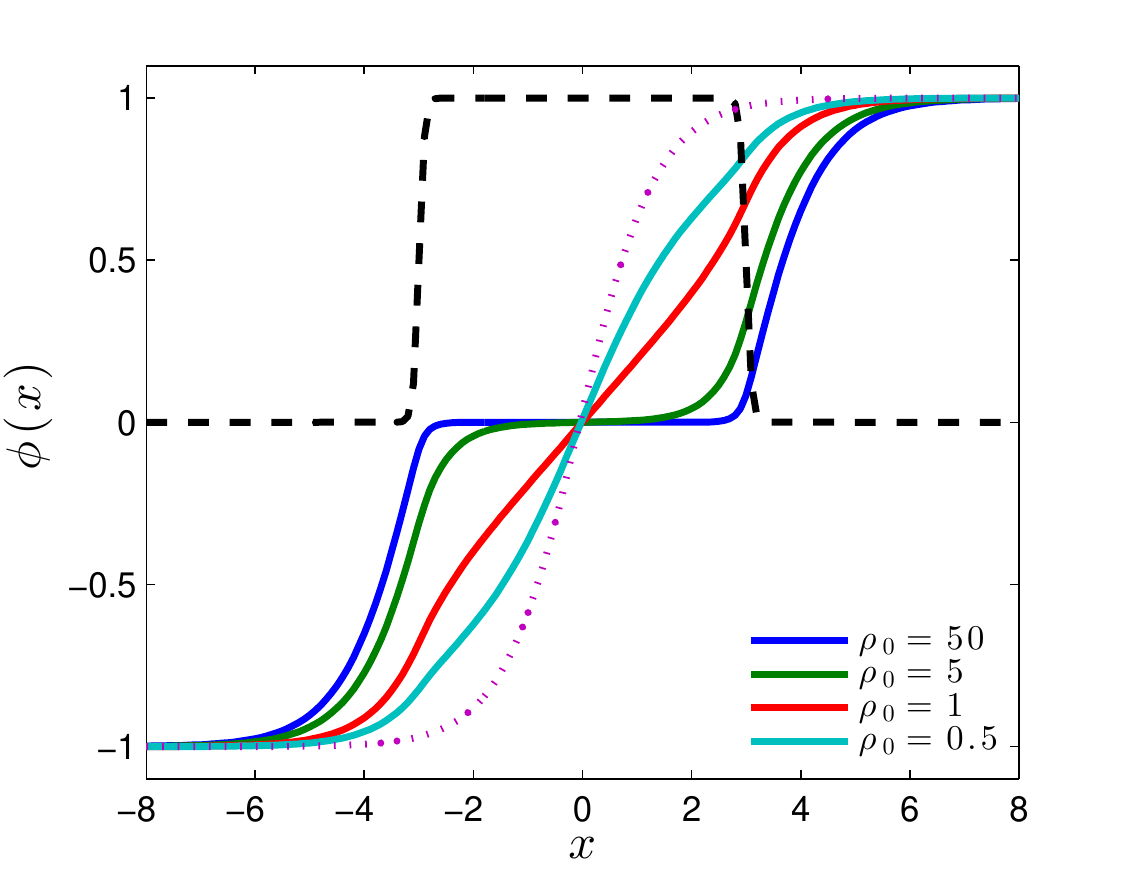}}	}
      	 \subfigure[$\, \mathcal{E}(x)$]{{\includegraphics[scale=0.6]{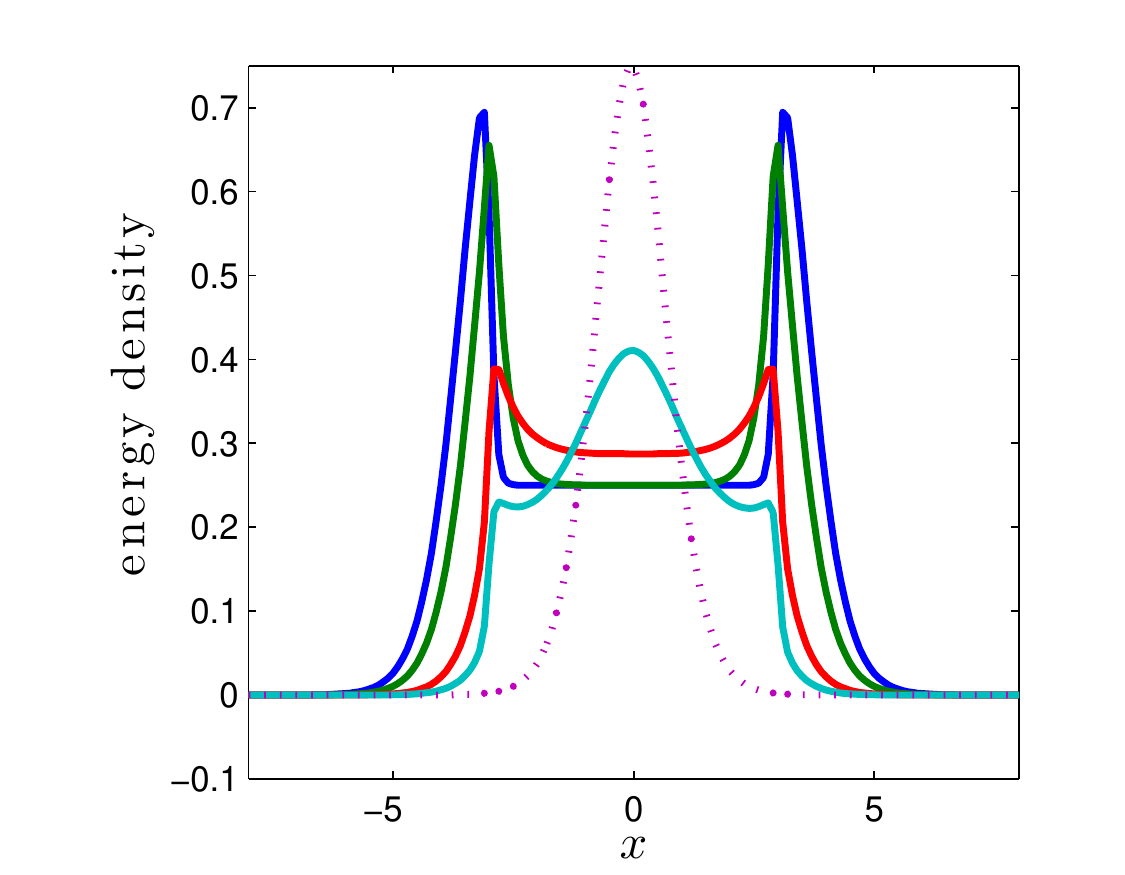}}	} 
      \end{center}
\caption{ Scalar field and energy density profiles of  1D static solutions of the symmetron equation with a ``tanh'' impurity density profile (\ref{eq:sec:1profile}). The   colours correspond to   different values of the internal density $\rho_0$ as shown in the legend in panel (a).  The dashed line in (a) corresponds to the shape of the impurity density profile, and the dotted line in both panels denotes the profiles in the absence of impurities.} \label{fig:1dsol}
\end{figure}

We will now present some 1D static solutions   obtained by gradient flow for an impurity density profile given by (\ref{eq:sec:1profile}). In \fref{fig:1dsol} we plot the static solution of the 1D equation of motion for a variety of values of $\rho_0$, with $r_0 = 3$ and $s = 0.1$. We set the boundary conditions
\bea
\phi(x = \pm\infty) = \pm 1.
\eea

One can observe the scalar field smoothly interpolating between the spatially separated vacua: this is the domain wall. For the systems with $\rho_0$ big enough (i.e., $\rho_0>1$) to restore the $\mathbb{Z}_2$ symmetry one can observe that inside the impurity, $|x|<r_0$, the scalar field profile is approximately zero;   there is a residual energy density inside the object   due to the  fact that $V(\phi=0) = \tfrac{1}{4}$. In the vicinity of the surface of the impurity there is a large energy density ``shell'': this comes from  the gradient contribution to the energy density $\mathcal{E}$. It is apparent that increasing $\rho_0$ has the effect of making the scalar field profile steeper at the boundary of the impurity, thereby decreasing the scalar field inside the impurity and further increasing the gradient energy density near the impurities surface, $|x|\approx r_0$. This is one of the reasons our numerical experiments are limited to certain ranges of values of $M$: when increasing $\rho_0$ these scalar field gradients quickly become hard to resolve on the lattice, introducing significant numerical uncertainties. For the small values of $\rho_0$ the scalar field  is only negligibly small at $x=0$ (at which point $\phi$ actually vanishes), and the energy density inside the impurity is dominated by the gradient contributions (which peak for the smallest values of $\rho_0$ at the origin).

\section{Idealized configurations}
\label{sec:idealized}
We would like to understand the  configurations of the symmetron system, and particularly those which have stable (or at least, metastable) domain walls. There are likely to be many interesting configurations one can make: we have constructed two types which we shall describe here. The idealized configurations we construct  will not precisely match onto objects found in realistic situations, but they will enhance   the understanding    of possible outcomes of evolving the equations of motion. We note that the existence of these configurations is only possible due to the symmetry-restoring impurities.  

\subsection{Domain wall pinning}
The first question we ask is: ``what happens if a domain wall is near an impurity?''. Our numerical answer to the question uses gradient flow to find  energy minimizing configurations. We   put down a single impurity, which has radius $r_0$, and place a single domain wall next to it. What happens next   will depend upon their separation. 

Before we continue we want to make our numerical methodology perfectly clear. For boundary conditions we make the $y$-direction periodic, and fix the values of the fields on the $x$-boundaries. This could have the effect of generating unphysical artifacts, but they are absent for times less than the ``light crossing time'', which is the amount of time taken for a signal travelling at the speed of light to traverse the periodic directions: all results we show are inside the light crossing time. We use gradient flow until $t=50$ in order to (a) smooth out the initial conditions, and (b) prolong the length of time the simulation can be run before hitting light-crossing issues.

\begin{figure*}[!t]
      \begin{center}
      	\subfigure[\, $t=0$ ]{\includegraphics[scale=0.15]{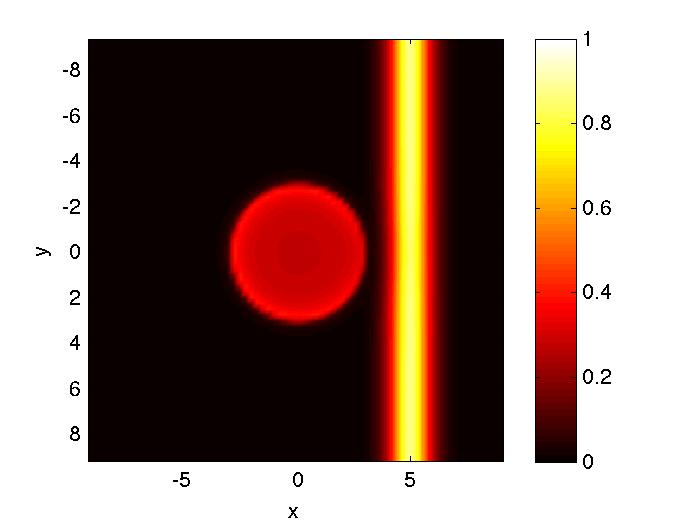}}
      	\subfigure[\, $t=10$ ]{\includegraphics[scale=0.15]{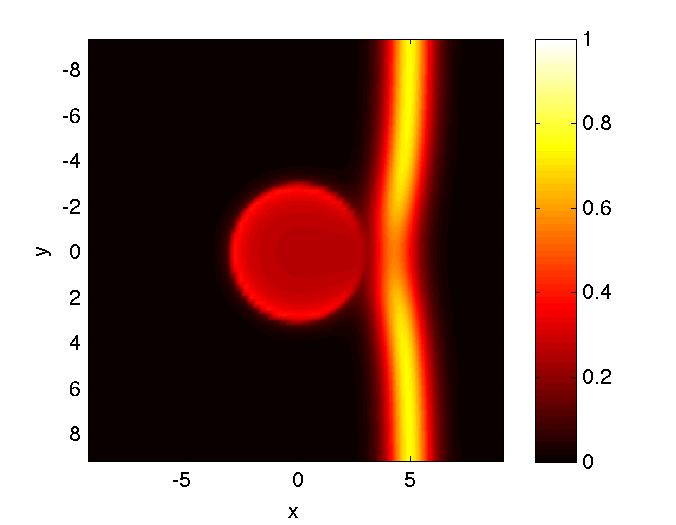}}
      	\subfigure[\, $t=20$ ]{\includegraphics[scale=0.15]{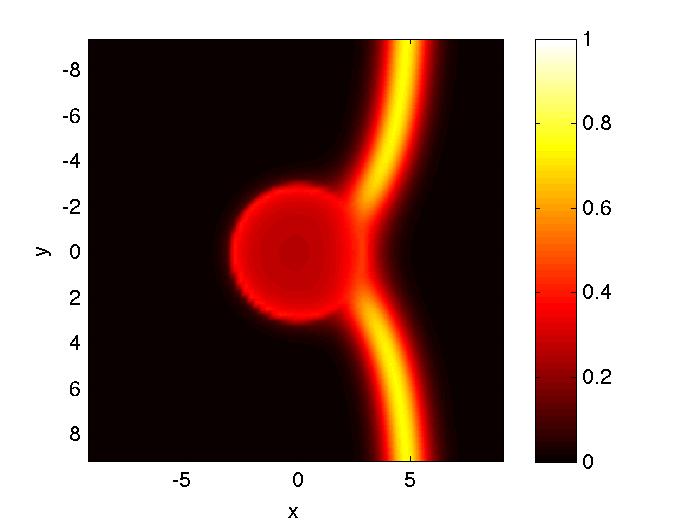}}
      	\subfigure[\, $t=30$ ]{\includegraphics[scale=0.15]{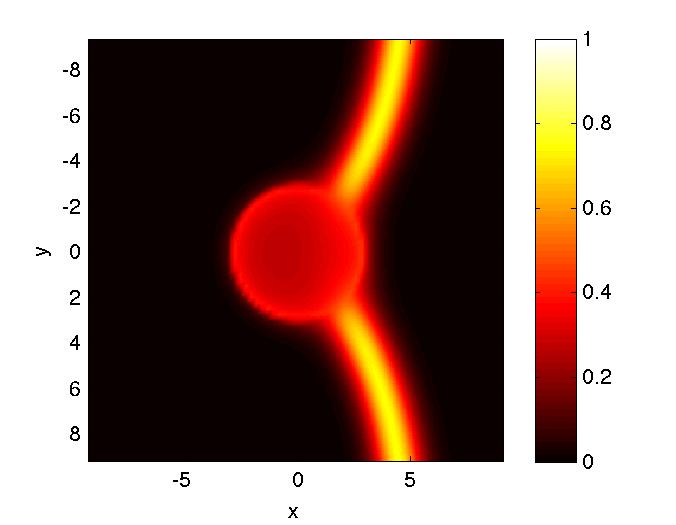}}
      	\subfigure[\, $t=40$ ]{\includegraphics[scale=0.15]{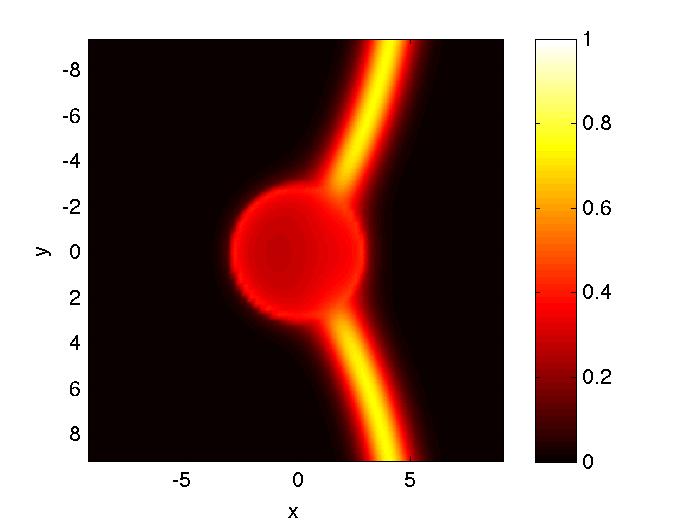}}				
      	\subfigure[\, $t=50$ ]{\includegraphics[scale=0.15]{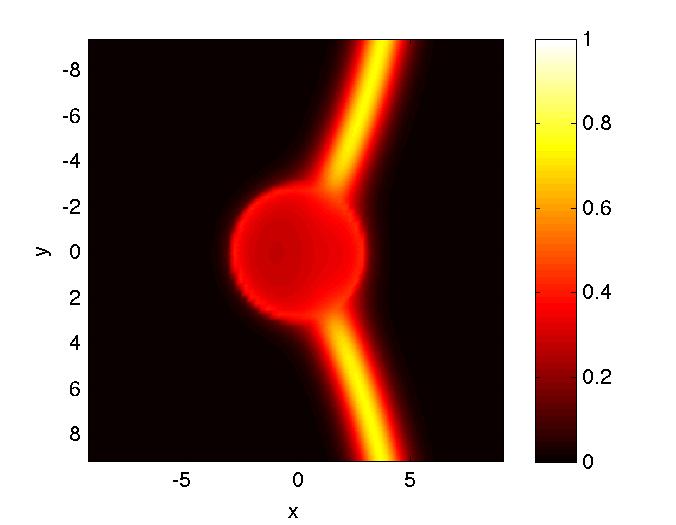}}				
      	\subfigure[\, $t=60$ ]{\includegraphics[scale=0.15]{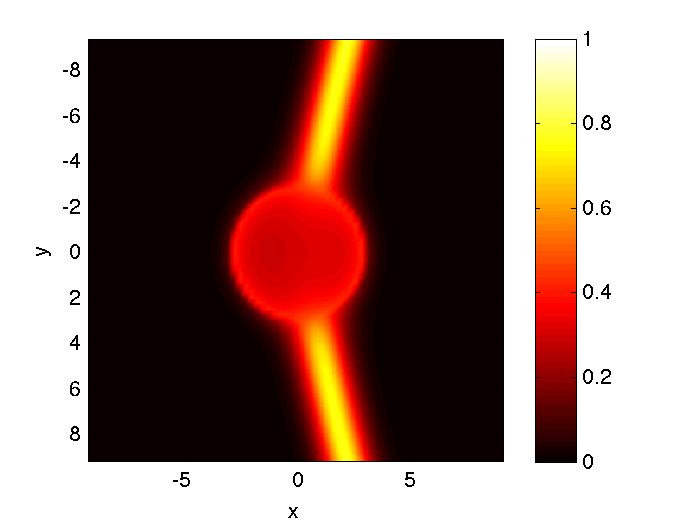}}				
      	\subfigure[\, $t=70$ ]{\includegraphics[scale=0.15]{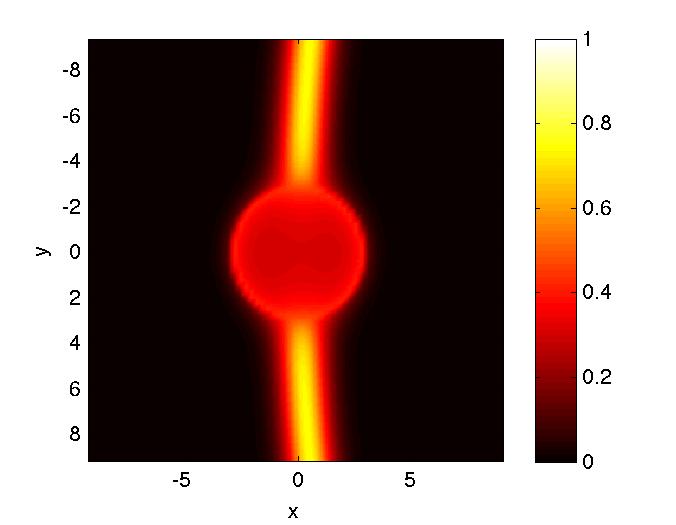}}				
      \end{center}
\caption{Evolution of the energy density for a scenario which illustrates the phenomenon of ``domain wall pinning''. The impurity   has radius $r_0=3$ and the domain wall is placed along the $y$-axis at $x_0 = r_0 +2$ (with zero velocity). It is clear that the domain wall is attracted by the impurity, and then the domain wall pinches off to pin to the distribution: these ends then travel towards the poles of the impurity. As explained in the main text, gradient flow is used until $t=50$, after which the   second order time derivative equation of motion is used; as such one shouldn't take the actual value of the ``times'' shown too seriously.} \label{fig:onewall_oneblob}
\end{figure*}

Putting the domain wall  down the $y$-axis at $x_0 = r_0 + 2$ gives rise to the following behavior. We find that the domain wall is attracted to the impurity; once the wall touches the outermost   point on the impurities equator, the wall breaks into two pieces. These pieces travel around the impurity until they reach the poles, at which point they stop. The portion of the domain wall which did not hit the impurities equator lags behind, but it eventually catches up, leaving a system where two domain walls terminate on opposite ends of the impurity. We give images of the evolution in \fref{fig:onewall_oneblob}.

We have therefore observed the phenomenon of domain wall pinning. The interpretation is that domain walls preferrentially terminate in high-density environments. This will have an impact on the evolution of the number density of  domain wall networks which form from phase transitions.  
 
\subsection{Domain wall necklace}
\label{sec:dwneck}
If one was to create a loop of domain wall with some radius, $R$ say, then it would collapse because each sequentially smaller loop has smaller length and therefore lower energy (see, e.g., \cite{Gregory:1989gg, Arodz:1993sy}). This picture can be altered by introducing   impurities into the path of a collapsing wall since we expect the walls to pin to impurities.

We will evenly distribute $N$ impurities on a circle of radius $R_0$. For simplicity we will study   the case in which each impurity has the same size properties (i.e., they   have the same radius $r_0$ and skin thickness $s$); if there are $N$ impurities then the coordinates of the centre of the $\qsuprm{n}{th}$ impurity is
\bea
\left(x_{0,n} ,y_{0,n}\right) = R_0 \bigg(\cos\left( 2\pi n/N\right),\sin\left(2\pi n /N\right) \bigg).
\eea
Given this collection of impurities we would like to know if there exists an energetically preferred configuration whereby domain walls link the impurities: this will form a \textit{domain wall necklace}. The reason why such a configuration could be interesting to find is that it has the potential to modify the scaling dynamics of domain wall networks. It is the phenomenon of domain wall pinning which fixes the walls onto the impurities.

The heuristic way to understand whether or not domain wall necklaces are possible   is quite simple:   compute the circumference of a large circle with pieces removed (the lengths of the removed pieces is determined by the sizes of the impurities) and compare that to the circumference of the largest internal circle which isn't punctured by the impurities. If the latter is larger than the former, then the energy minimizing configuration can be expected to be a sequence of domain walls which connect the impurities in a necklace-like configuration. There is another consideration to be taken into account: the impurities should not touch. 

There are configurations with $N\geq 4$ impurities which can satisfy both of these conditions -- the calculation we are perform is not sufficiently sophisticated to concretely predict a complete list of allowed configuration. However, it does allow us to explain why the $N=3$ case is qualitatively different.

\subsubsection{Calculation of the bound in the $N=3$ case}
Our aim is to understand whether or not a configuration with $N=3$ impurities can form a domain wall necklace. As explained above, a necklace can only be expected to exist if (a) it is of preffered energy, and (b) the impurities do not touch. Here we compute both of these ``bounds'' on the configuration of the impurities, and ultimately show that $N=3$ is not capable of simultaneously satisfying both. We will work with a general value of $N$.

We first estimate the properties of a configuration for whom a domain wall necklace is an energetically preferred state. To do this we estimate the energy of the domain wall necklace, and compare it to the energy of the domain wall  which lives inside the necklace. This requires the length of domain wall which makes up the necklace: this is a simple but tedious geometrical exercise.

\begin{figure}[!t]
      \begin{center}
       {\includegraphics[scale=0.4]{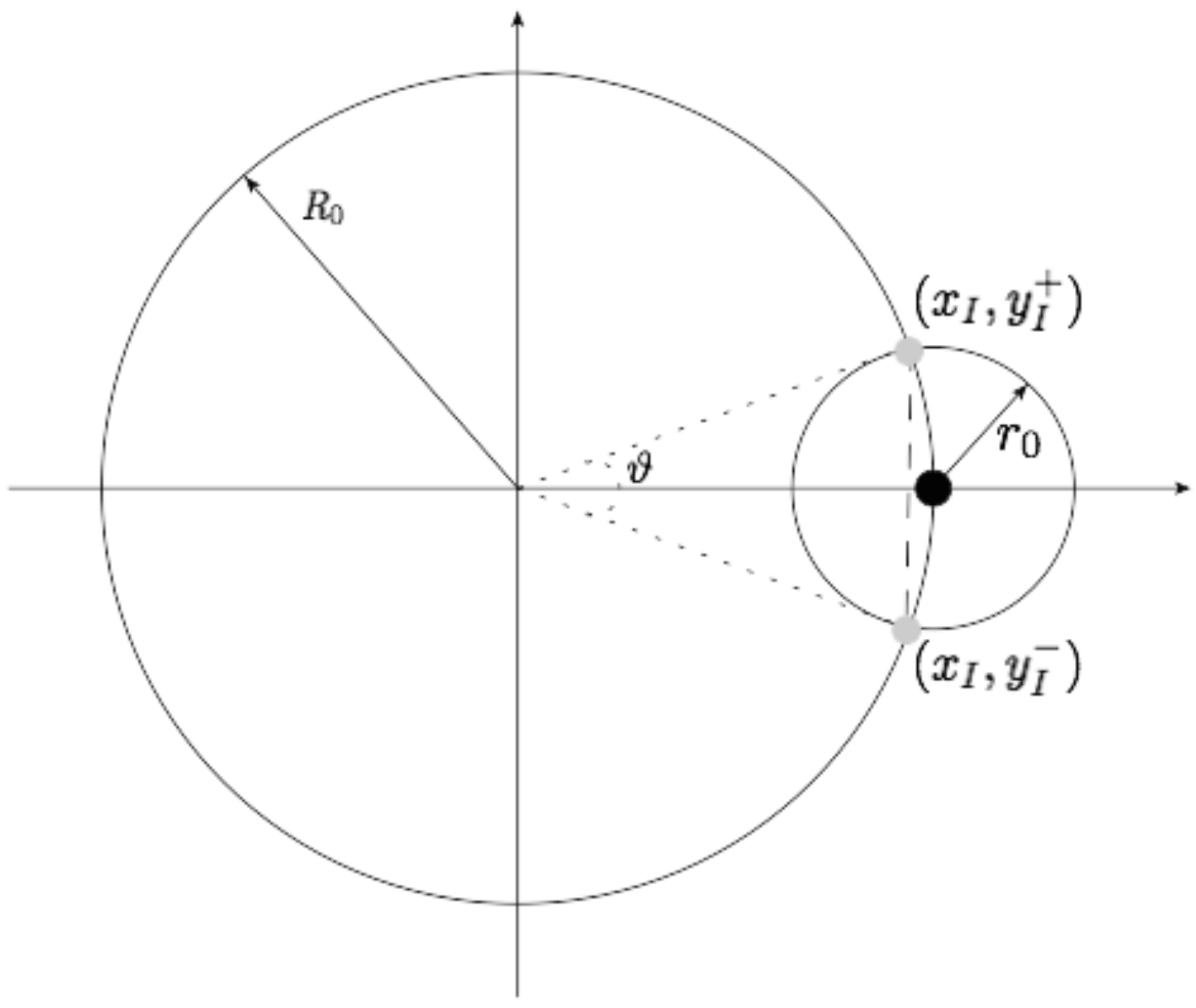}}
      \end{center}
\caption{  Steps to constructing the energetic bound for the domain wall necklace. The impurities are the small circles of radius $r_0$ that live on the large circle of radius $R_0$. The coordinates of intersection are $(x_I,y_I^{\pm})$, and the angle subtended by the arc of the large circle through them is $\vartheta$. } \label{fig:necklace_constrict}
\end{figure}

The construction we are about to describe is depicted in \fref{fig:necklace_constrict}.
Consider two circles: one centred on the origin with radius $R_0$ and a second whose centre has coordinates $(R_0,0)$ and has radius $r_0$. The equations of the circles are given by
\bse
\label{eq:sec:circles-1}
\bea
\label{eq:sec:circles-1-a}
x^2 + y^2 = R_0^2,
\eea
\bea
\label{eq:sec:circles-1-b}
\left(x-R_0\right)^2 + y^2 = r_0^2.
\eea
\ese
We are interested in obtaining the coordinates of intersection of these two circles. That is, we want to find the set of coordinates $\{(x_I,y_I)\}$  that are common to both circles, and are found by solving (\ref{eq:sec:circles-1}). After subtracting   (\ref{eq:sec:circles-1-b}) from (\ref{eq:sec:circles-1-a}), it follows by simple  rearrangement that
\bea
\label{eq:sec:circles-1-a-dd}
x_I = R_0 - r_0\left(\frac{r_0}{2R_0}\right).
\eea
Putting the $x$-coordinate of intersection (\ref{eq:sec:circles-1-a-dd}) into  (\ref{eq:sec:circles-1-b}) we obtain the $y$-coordinates of intersection,
\bea
y_I^{\pm} = \pm r_0\sqrt{1-\left( \frac{r_0}{2R_0}\right)^2}.
\eea
Now imagine drawing straight lines from the origin to the points of intersection; $(0,0)\rightarrow (x_I,y_I^+)$ and $(0,0)\rightarrow (x_I,y_I^-)$. The angle subtended by these lines, $\vartheta$, is given by
\bea
\vartheta = 2\tan^{-1}\left( y_I^+/x_I\right).
\eea
Hence, the length of the arc of the ``large'' circle (which was centred on the origin and had radius $R_0$) which lies within the smaller circle is $\qsubrm{l}{arc} = R_0\vartheta$.
If there are $N$   such circles (each with radius $r_0$, and each being centred on the large circle) then the total length of the arcs inside these small circles is
\bea
\qsubrm{l}{REM} = N\qsubrm{l}{arc}.
\eea
This is the amount of the   total circumference  of the circle which is to be removed; the total circumference is (trivially) $\qsubrm{l}{TOT} = 2\pi R_0$.
And so, the length of the large circle which is occupiable by a domain wall is
\bea
\qsubrm{l}{occ} = \qsubrm{l}{TOT} - \qsubrm{l}{REM} = 2\pi R_0\left( 1 - \frac{N\vartheta}{2\pi}\right).
\eea
The biggest circle inside $R_0$ that isn't punctured by the impurities begins at the inside edge of the impurities and whose circumference is
\bea
\qsubrm{l}{IN} = 2\pi(R_0 - r_0).
\eea
When $\qsubrm{l}{IN}$ is bigger than $\qsubrm{l}{occ}$ (which was $\qsubrm{l}{TOT}$ with the length $\qsubrm{l}{REM}$ removed by the puncturing of the impurities), then the energy minimizing loop will be a set of links joining the impurities -- forming a ``necklace''.  It should be apparent that  $R_0 > \half r_0$ is required for $\qsubrm{l}{REM}$ to return a real result.

We now use this to estimate the energy of the domain walls ignoring the effects of curvature. The energy of a circular wall of radius $r$ is $E \sim 2\pi r \sigma$, where $\sigma$ is the wall's energy per unit length. The estimate of the energies of walls whose radii $\qsubrm{r}{w}$ are much larger than $R_0$, equal to $R_0$, and much less than $R_0$, respectively are
\bse
\bea
E(\qsubrm{r}{w}\gg R_0) \sim 2\pi \qsubrm{r}{w}\sigma,
\eea
\bea
E(\qsubrm{r}{w}=R_0) \sim 2\pi R_0\sigma\left( 1 - \frac{N\vartheta}{2\pi}\right),
\eea
\bea
E(\qsubrm{r}{w}\ll R_0) \sim 2\pi \qsubrm{r}{w}\sigma.
\eea
\ese
If a configuration exists for whom
\bea
\label{bounda_energy}
E(\qsubrm{r}{w}=R_0) < E(\qsubrm{r}{w}\ll R_0)
\eea
then it follows that the configuration at radius $\qsubrm{r}{w}=R_0$ is energetically preferred over the one at some radius $\qsubrm{r}{w}< R_0$ (it can't be expected to be preferred for all $\qsubrm{r}{w}\ll R_0$).

The additional required constraint is that the impurities should not touch. If they were to touch, a domain wall could never collapse to the interior, negating the interest of the configuration. To compute this constraint, one should imagine a modification to \fref{fig:necklace_constrict}, whereby another impurity of radius $r_0$ is placed at coordinates
\bea
R_0\left(\cos {2\pi}/{N},\sin {2\pi}/{N}\right).
\eea
By drawing a line through the centres of this new impurity, and the one already present, one obtains the following condition that the two should not touch:
\bea
\label{eq:sec:notouchcondition}
\frac{R_0}{2r_0}\left( 1 + \cos\frac{2\pi}{N}\right)>1.
\eea 
One can verify that there are no configurations with $N=3$ impurities which simultaneously satisfy   the preferential-energy (\ref{bounda_energy}) and the no-touch (\ref{eq:sec:notouchcondition}) conditions. Therefore, we should not expect $N=3$ necklaces to exist, at least on the basis of these simple calculations.

\begin{figure}[!t]
      \begin{center}
      \subfigure[\, $N=4$ ]{\includegraphics[scale=0.4]{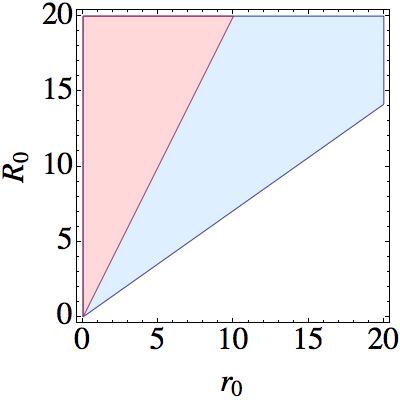}}       \qquad
            \subfigure[\, $N=7$ ]{\includegraphics[scale=0.4]{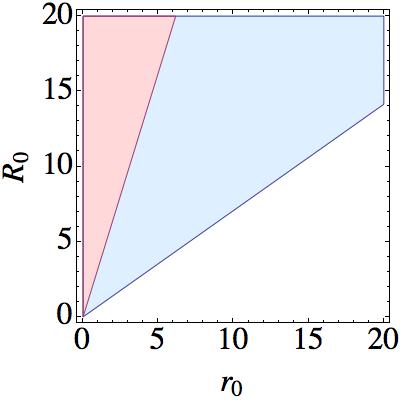}}      
      \end{center}
\caption{Plots showing the anecdotal conditions for the allowed configurations   of domain wall necklaces. We remind that there are $N$ impurities, each of size $r_0$, evenly distributed on a circle of radius $R_0$. The blue shaded region shows where the preferred energy  condition (\ref{bounda_energy}) holds on the circle of radius  $\qsubrm{r}{w} = R_0-r_0$, and the red shaded region shows where both the preferred energy    (\ref{bounda_energy}) and no-touch conditions (\ref{eq:sec:notouchcondition}) hold. There is no overlap between the regions in the equivalent plots for $N=3$.  } \label{fig:necklace_existence}
\end{figure}

In \fref{fig:necklace_existence} we plot the regions in the $(r_0,R_0)$-plane for two values of $N$ where the preferential-energy-condition (\ref{bounda_energy}),  on the maximal-internal circle with radius $\qsubrm{r}{w} = R_0-r_0$, and  no-touch-condition (\ref{eq:sec:notouchcondition}) are satisfied. We should note that this very simple calculation only provides anecdotal evidence for the existence of domain wall necklaces. In the next section we give numerical evidence for their existence.

\subsubsection{Numerical construction}
We now present some domain wall necklace solutions, which we construct in the following manner.
We place a set of impurities in a desired configuration, and then setup a domain wall encompassing the impurities. The gradient flow equations then cause the domain wall loop to shrink until an energy minimizing configuration is obtained. There are $N$ equally spaced impurities, each with radius $r_0 = 3$, that sit on a circle of radius $R_0 = 10$. The impurities have central density $\rho_0 = 5$. In our numerics we use time-step size $\Delta t = 0.01$,  space step-size $\Delta x = 0.25$; we use fixed boundary conditions, but there isn't the issue of unphysical artifacts in this case.

\begin{figure*}[!t]
      \begin{center}
      	\subfigure[\, $t=0$ ]{\includegraphics[scale=0.2]{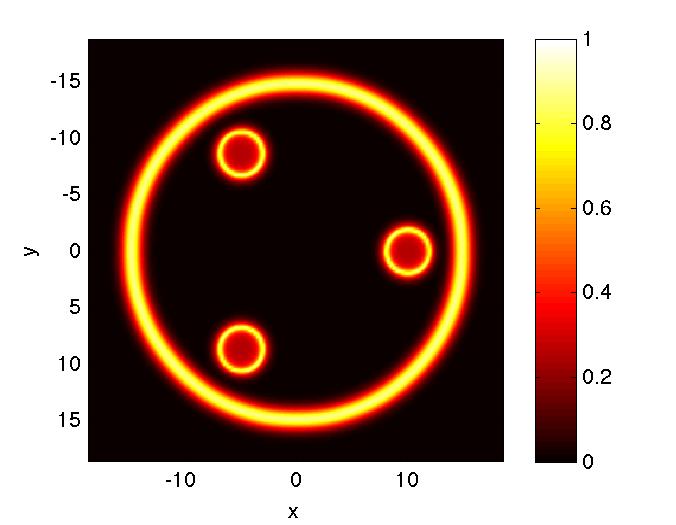}}
	\subfigure[\, $t=20$ ]{\includegraphics[scale=0.2]{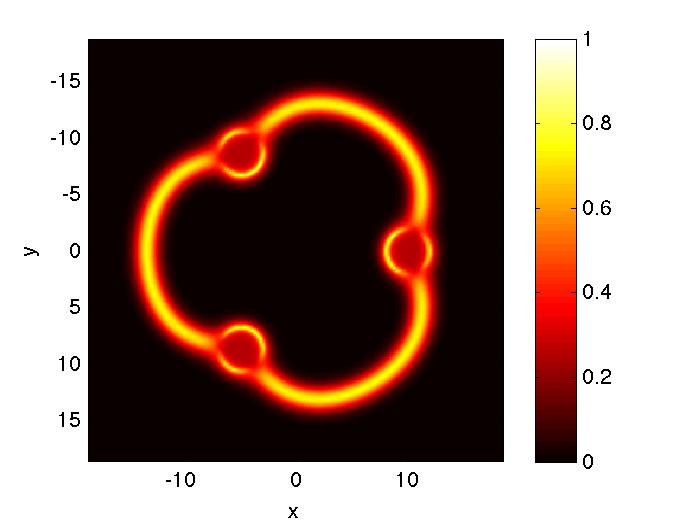}}      
	\subfigure[\, $t=40$ ]{\includegraphics[scale=0.2]{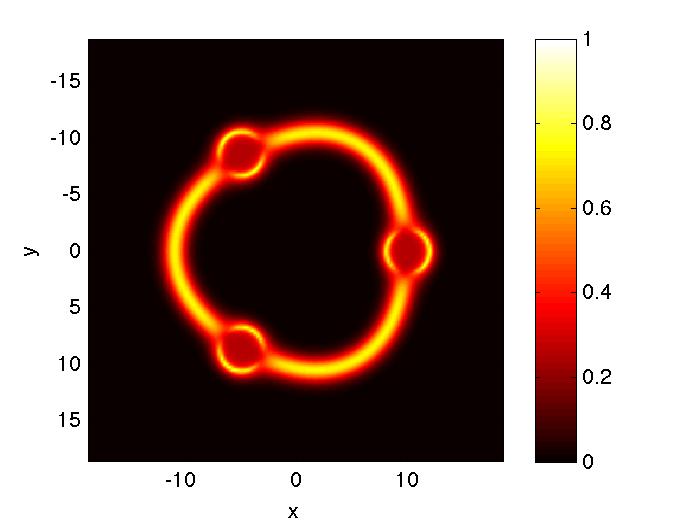}}	
	\subfigure[\, $t=120$ ]{\includegraphics[scale=0.2]{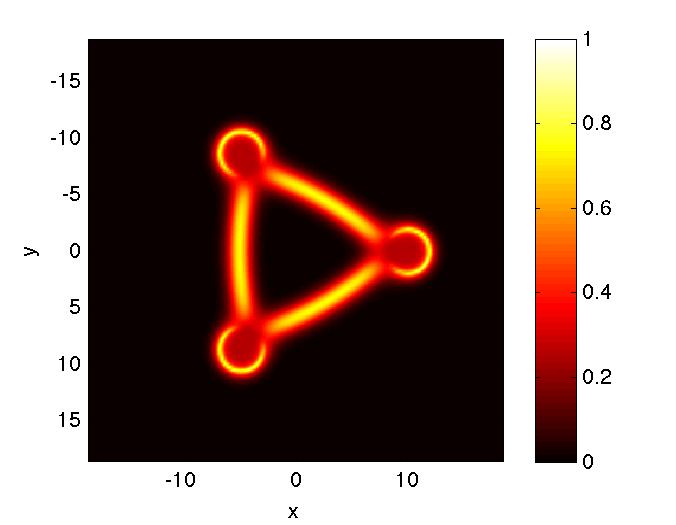}}	
	\subfigure[\, $t=130$ ]{\includegraphics[scale=0.2]{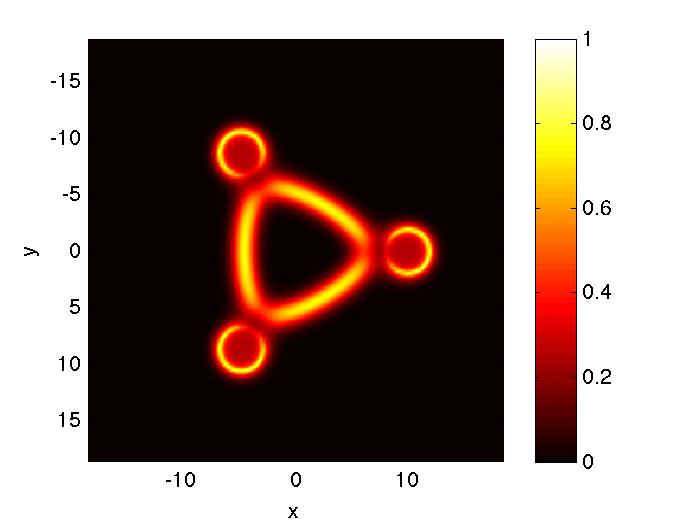}}		
	\subfigure[\, $t=140$ ]{\includegraphics[scale=0.2]{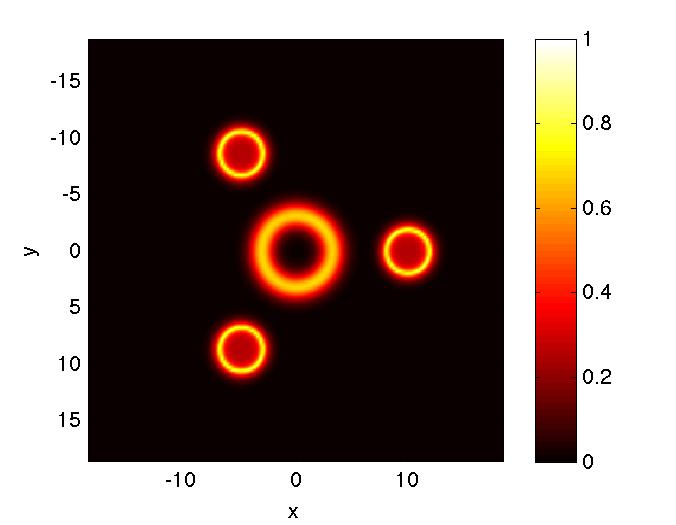}}		
      \end{center}
\caption{ Evolution of the energy density during an attempt to construct an $N=3$ domain wall necklace. it is clear that the domain wall loop collapses after it reconnects in the interior of the circle which the impurities sit on.  } \label{fig:necklace:N3}
\end{figure*}

 In \fref{fig:necklace:N3} we plot an attempt at constructing an $N=3$ domain wall necklace. It is apparent that the attempt has failed since after the walls pin to the impurities, the   walls  re-connect in the interior of the circle on which the impurities live, forming a loop which collapses in the standard manner. This is in accord with the remarks we made at the end of the previous subsection.

\begin{figure*}[!t]
      \begin{center}
      	\subfigure[\, $t=0$ ]{\includegraphics[scale=0.2]{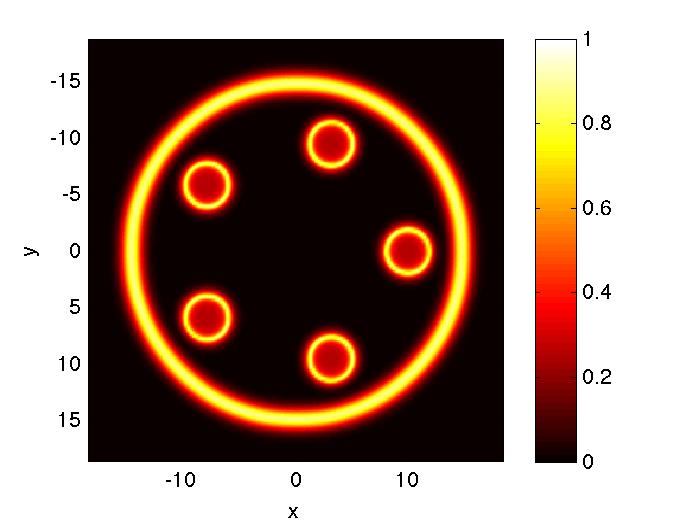}}
	\subfigure[\, $t=20$ ]{\includegraphics[scale=0.2]{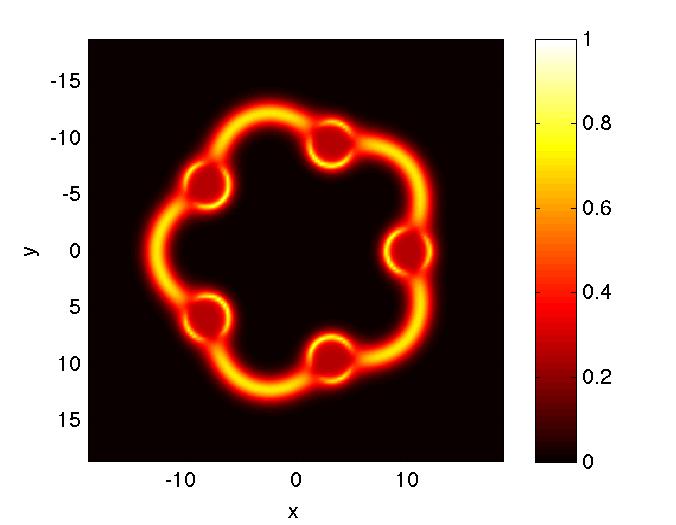}}      
	\subfigure[\, $t=40$ ]{\includegraphics[scale=0.2]{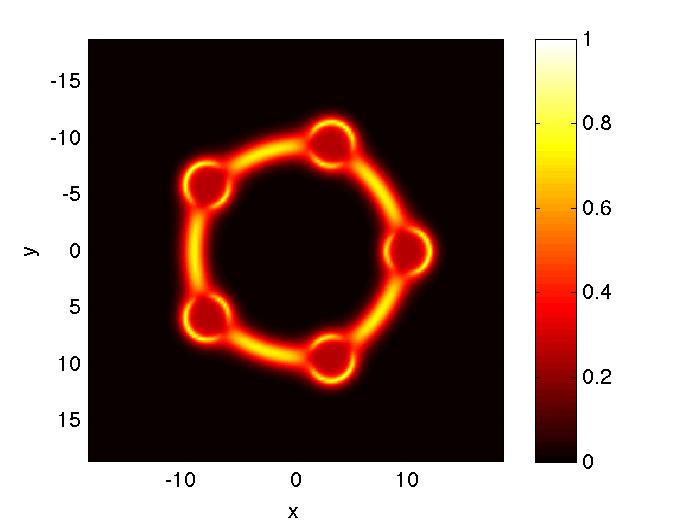}}	
	\subfigure[\, $t=120$ ]{\includegraphics[scale=0.2]{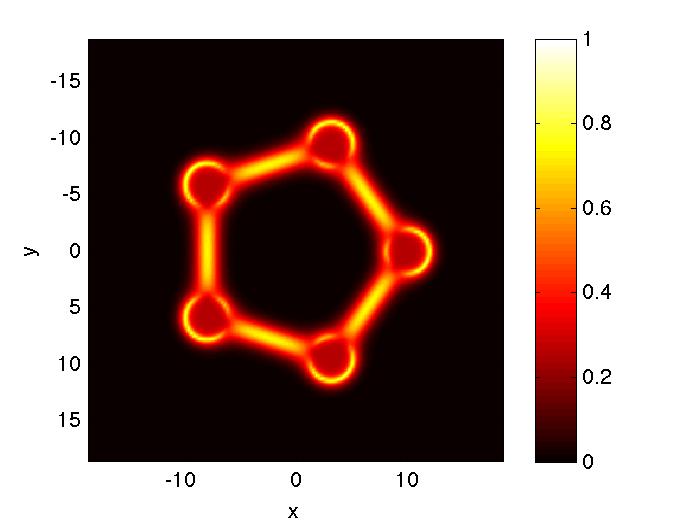}}	
	\subfigure[\, $t=130$ ]{\includegraphics[scale=0.2]{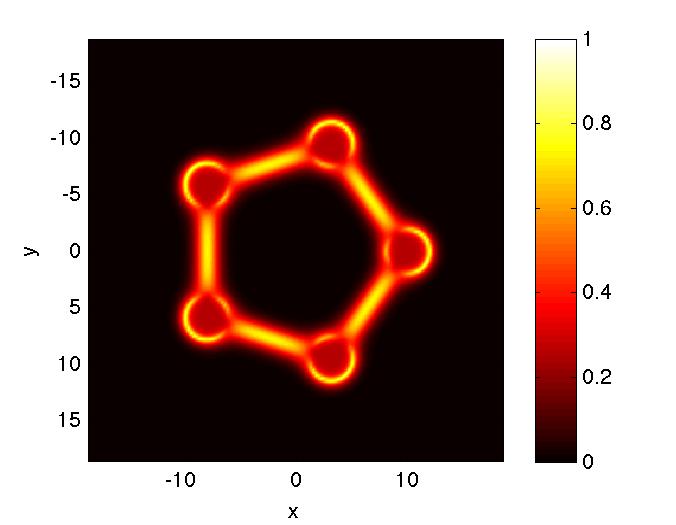}}		
	\subfigure[\, $t=140$ ]{\includegraphics[scale=0.2]{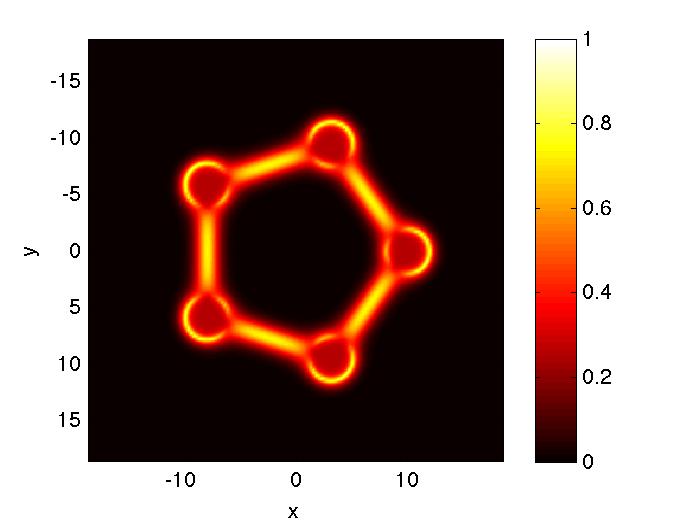}}	
      \end{center}
\caption{Finding an $N=5$ necklace solution: this is a plot of the energy density. It is clear that the initial domain wall loop collapses until it connects onto the impurities, after which the wall straightens out and stays connected to the impurities. } \label{fig:necklace:N5}
\end{figure*}

\begin{figure*}[!t]
      \begin{center}
      	\subfigure[\, $t=0$ ]{\includegraphics[scale=0.2]{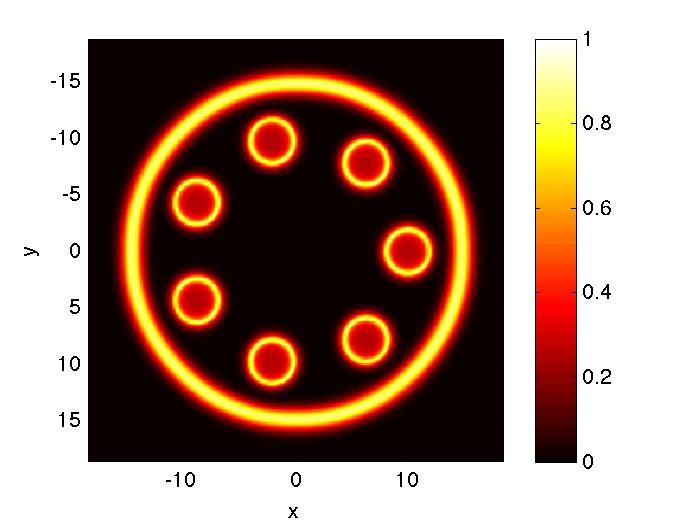}}
	\subfigure[\, $t=20$ ]{\includegraphics[scale=0.2]{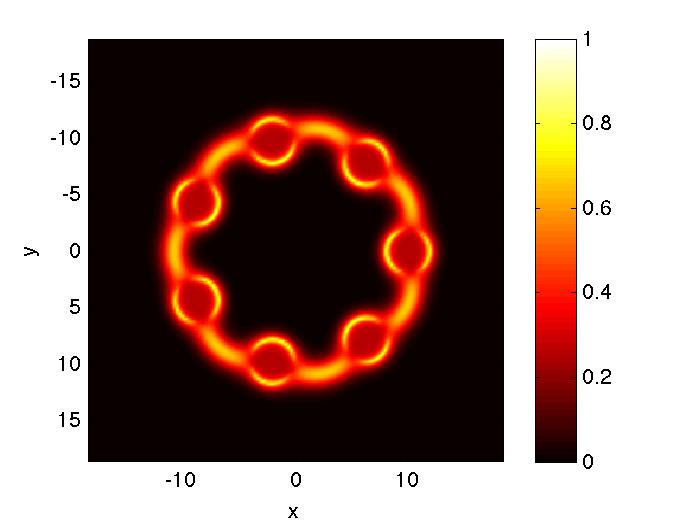}}      		
	\subfigure[\, $t=140$ ]{\includegraphics[scale=0.2]{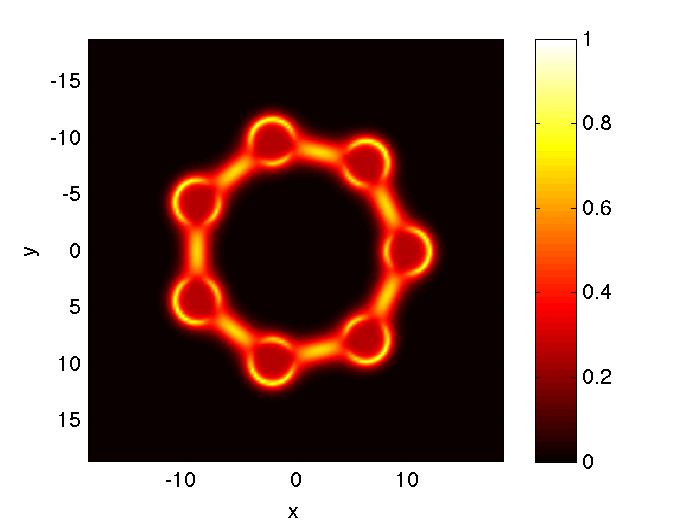}}		
      \end{center}
\caption{Evolution of the energy density during the construction of an $N=7$ domain wall necklace. We have skipped out images of the intermediate  time-steps for brevity. } \label{fig:necklace:N7}
\end{figure*}

In \fref{fig:necklace:N5} we plot the energy density at various times during the gradient flow algorithm, with $N=5$ impurities. There it is clear that we have constructed a metastable domain wall necklace. In \fref{fig:necklace:N7} we plot similar images but for $N=7$. It is interesting to note that the walls which link the impurities are (at least by eye) straight.

We would like to make it perfectly clear that   these configurations are at best meta-stable: if the loop were to contract further (i.e., below some radius which is well within the radius $R_0-r_0$, which could conceivably occur in a realistic scenario if some radiation effect pushed a portion the necklace inwards) then the energy minimizing configuration would just be ``no loop''.  In the next section we will take a look at the chances of necklaces forming from random initial conditions.

\section{Dynamical random networks}
\label{eq:sec;dyn-rand}
We now move on to study the dynamics of more ``realistic'' systems, in the sense that domain walls evolve from random initial conditions. The systems will have a set of $N$ impurities, specified by the profiles discussed in section \ref{sec:impurity-prfiles}.

A few words on our numerical methodology. What we want to do is to evolve a network of domain walls (in the presence of impurities). To obtain such a network, we   set each lattice site in one of the two vacua randomly. This introduces huge and unphysical discontinuities into the field configuration (spatial gradients are infinite since the configuration is not smooth). To ameliorate this we modify the equation of motion for times $t < \qsubrm{t}{D}$, where $\qsubrm{t}{D}$ is a time which defines the cut-off for damped dynamics; this is the same strategy successfully used in \cite{Battye:2006pf, Battye:2009ac, Battye:2010dk, Battye:2011ff}. This process yields a smooth network of domain walls which evolve according to their relativistic equation of motion after the ``damping'' has been switched off. To summarise, we evolve the equation of motion
\bea
\label{eom-damoubg}
\ddot{\phi} + \alpha(t)\dot{\phi} = \nabla^2\phi - \frac{\dd V}{\dd\phi},
\eea  
where the damping coefficient is given by
\bea
\alpha(t) = \left\{ \begin{array}{cc} \alpha_0 & t < \qsubrm{t}{D},\\ 0 & t \geq \qsubrm{t}{D}.\end{array}\right.
\eea
Our simulations use space step-size $\Delta x = 0.25$, time step-size $\Delta t = 0.1$, and the physical length $L$ of the box is given in terms of the number of grid-points in each direction $P$ by $L = P\Delta x$. From hereon we only quote physical length and physical time. We use $ \qsubrm{t}{D}=20, \alpha_0=5$.

We will evolve the equation of motion for three distinct purposes. First, we want to study the scaling dynamics, focussing on how the presence of impurities affects the $\qsubrm{N}{dw}\propto t^{-1}$ scaling ``law''. Secondly, we want to understand what happens ``inside'' an impurity.  Our final purpose is to find out whether domain wall necklaces can be formed from random initial conditions. 

\subsection{Scaling dynamics}
To study the scaling dynamics of domain wall networks  requires a computation of the number of domain walls, $\qsubrm{N}{dw}$, in the simulation every time-step. This is done by checking whether the  field changes sign between a given lattice site and its upper and right-neighbours. Such simulations are   only physically meaningful for times $\qsubrm{t}{D}\leq t \leq \qsubrm{t}{lx}$, where $\qsubrm{t}{lx}$ is the light-crossing time, as discussed in the previous sections. We will use lattices with ``large'' numbers of grid-points, and a number of realizations to minimize  the effects of artifacts imposed from the random initial conditions.

\begin{figure}[!t]
      \begin{center}
 	 \subfigure[\,$L=256 \,(P = 1024)$]{\includegraphics[scale=0.63]{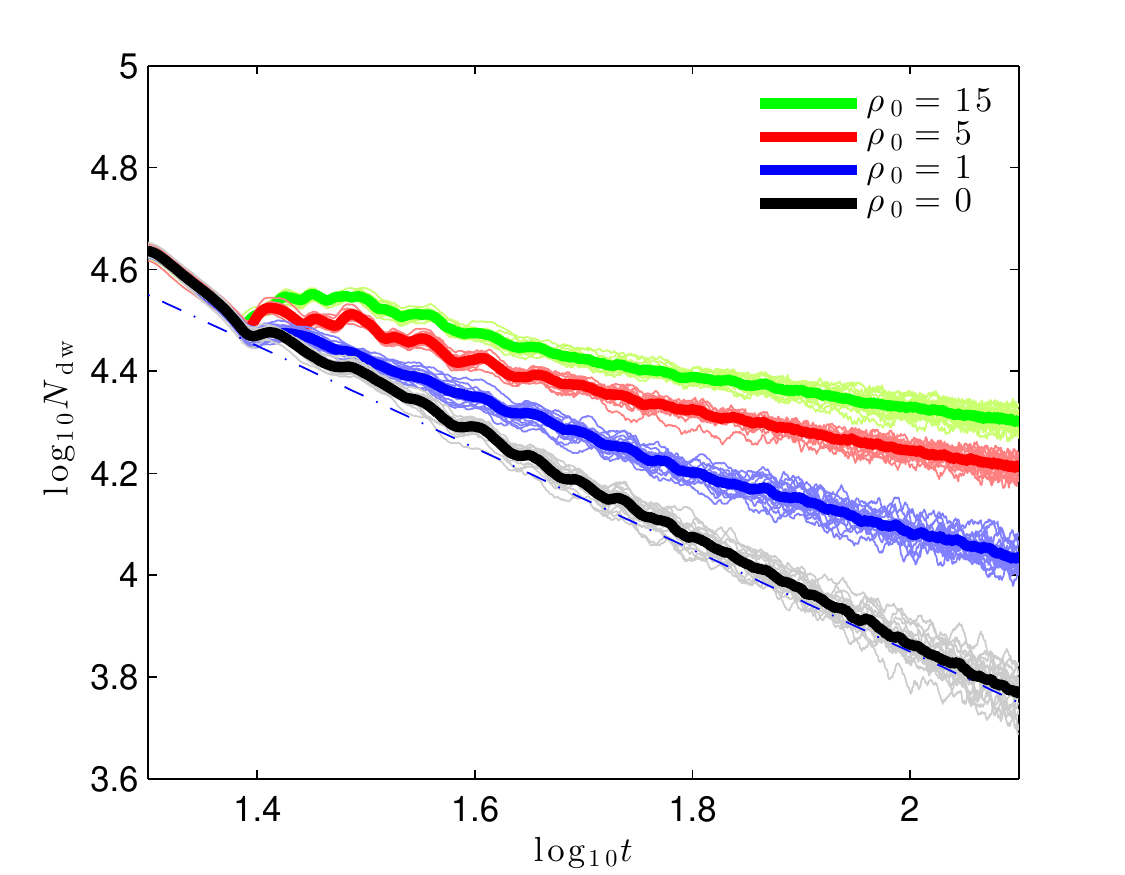}}
	 \subfigure[\,$L=512 \,(P = 2048)$]{\includegraphics[scale=0.63]{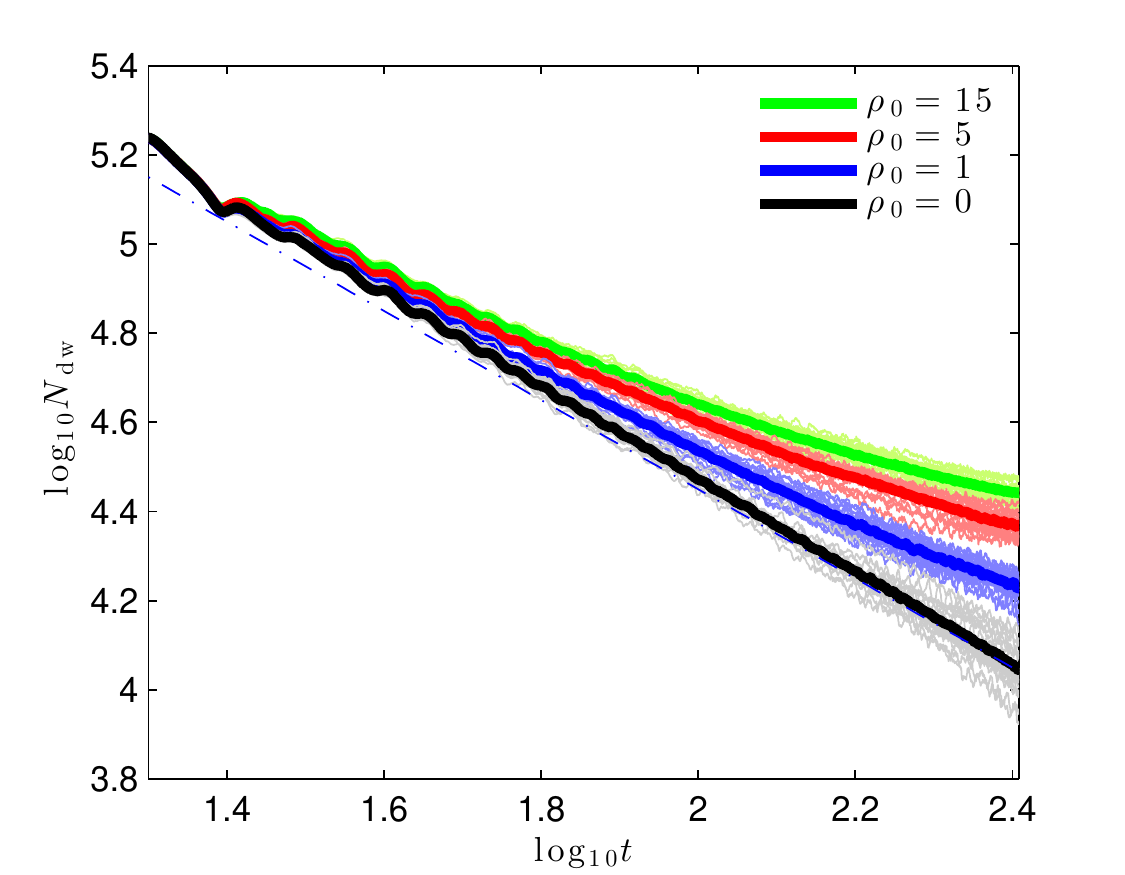}}
      \end{center}
\caption{Evolution of the number of domain walls from random initial conditions in the presence of impurities for two different box sizes. The physical lengths of the boxes, $L$, are shown in the captions, and for clarity we have given the number of grid-points in each direction. $P$.  These systems have $N = 10$ impurities, each with $r_0=10,s=0.1$,   equally spaced on a circle of radius $R_0 = 50$. For each physical length of the box there are four sets of simulations, each with a different value of the internal density $\rho_0$ as shown in the legend. For a given value of $\rho_0$ we have performed 20 realizations, as shown by the thin lines: the solid lines are the averages over the realizations. The dot-dashed line is the expected $\qsubrm{N}{dw}\propto t^{-1}$ scaling law (the system with $\rho_0=0$ agrees with this scaling law).} \label{fig:nwalls}
\end{figure}

In \fref{fig:nwalls} we plot the evolution of the number of domain walls from random initial conditions for systems containing $N=10$ impurities, on boxes with length $L=256$ and $L=512$. The impurities are evenly spaced on a circle of radius $R_0 = 50$,   each having   size $r_0 = 10$ and skin thickness $s=0.1$. Each thick line in the plot is an average over 20 realizations (the thin lines provide the evolution for each realization), and each colour corresponds to a different value of the internal density $\rho_0$ of the impurities. 

\begin{figure*}[!t]
      \begin{center}
 	 \subfigure[\,$L=256\, (P = 1024)$]{\includegraphics[scale=0.43]{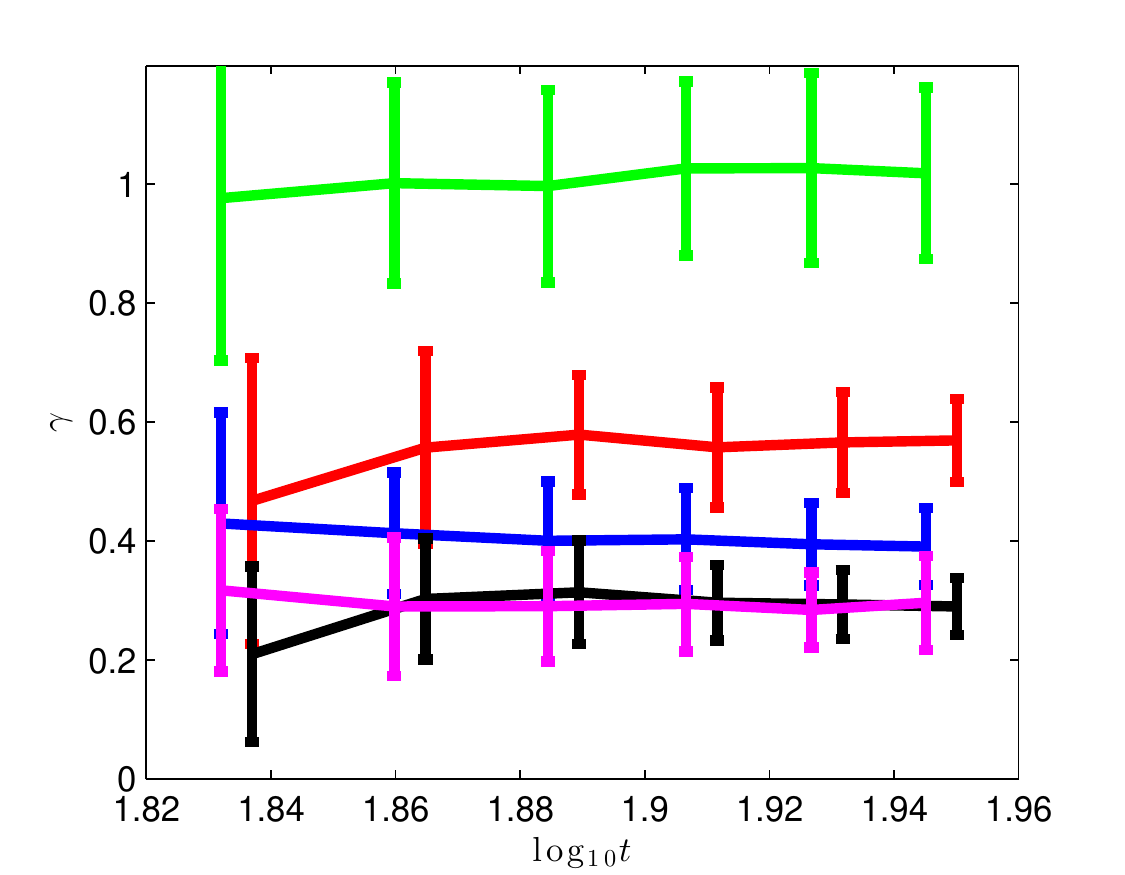}}
 	 \subfigure[\,$L=512\, (P = 2048)$]{\includegraphics[scale=0.43]{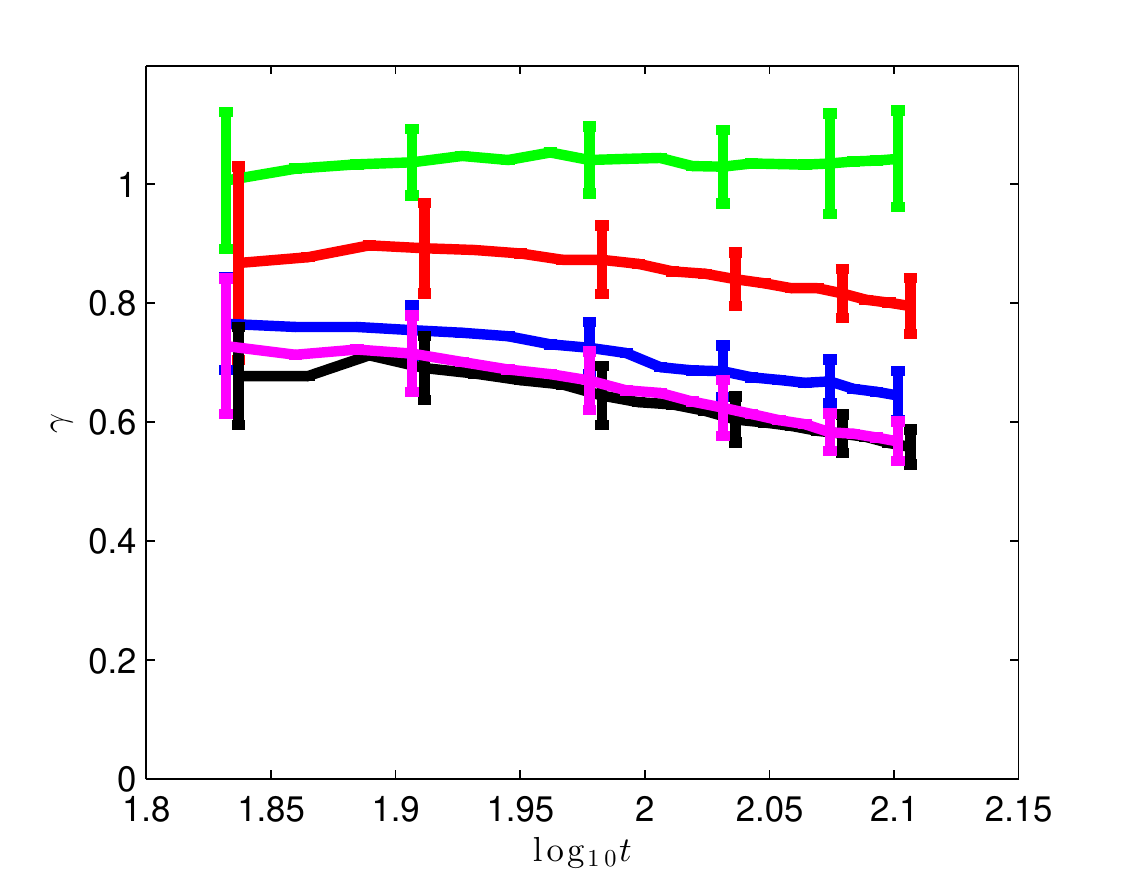}}
  	 \subfigure[\,$L=1024\, (P = 4096)$]{\includegraphics[scale=0.43]{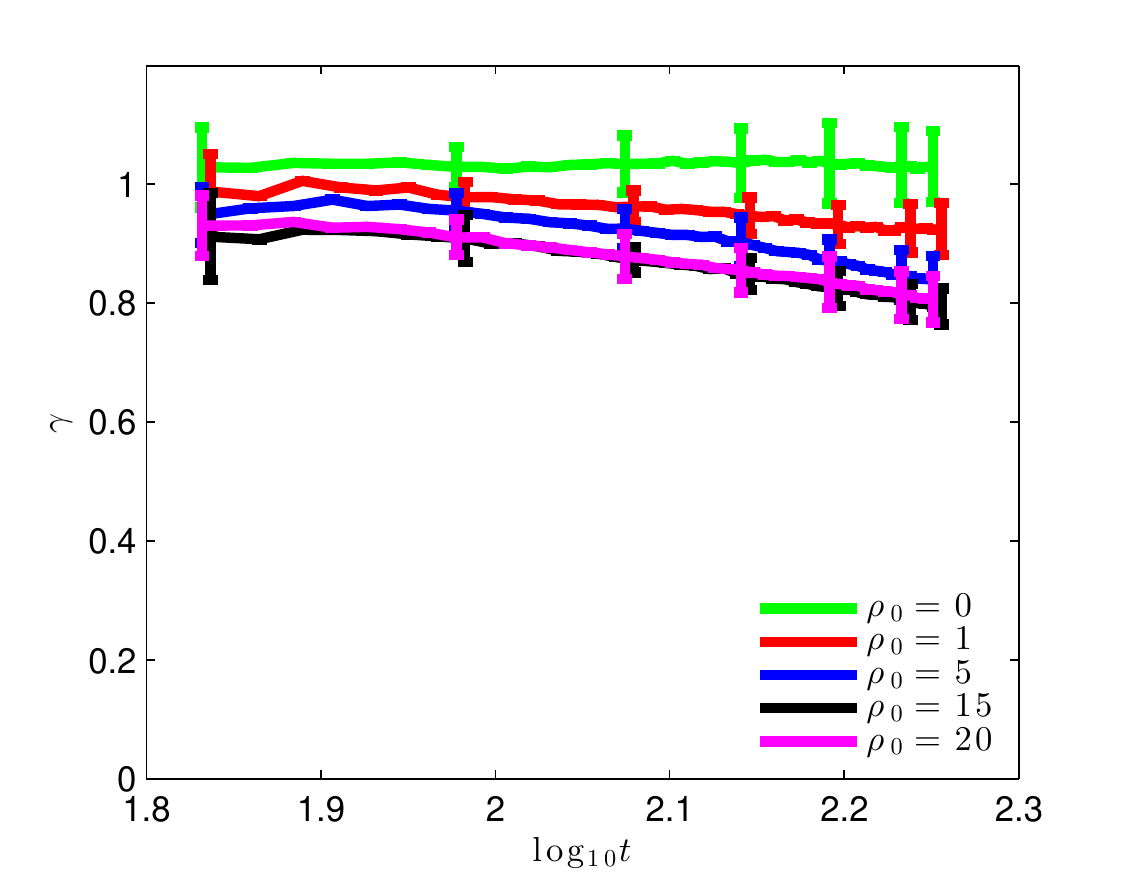}}
      \end{center}
\caption{ Evolution of the scaling exponents for the same distribution of impurities described in the caption of \fref{fig:nwalls}, with $1\sigma$-error bars and the results for a larger box included for completeness. To avoid cluttering the plots, the horizontal position of every-other line is shifted slightly to the right to aid viewing, as well as skipping out some error bars. We have computed the average of the evolution of the number of domain walls from 20 realizations; the scaling exponent $\gamma$ is computed from the average via $\overline{{N}}\qsubrm{}{dw}\propto t^{-\gamma}$, in bins of width $\qsubrm{\delta t}{bin} = 10$. The different colours corresponds to different values of the internal densities, as shown in the legend in the third panel (the colour schemes are consistent across each of the three panels). It is clear that increasing $\rho_0$ has quite a significant effect on the scaling exponent, as well as inducing a rather substantial evolution when the box size is increased. These plots also show that the critical exponent is reached by $\rho_0 = 15$.} \label{fig:gammaevol}
\end{figure*}

The figure shows that as the internal density is increased the scaling of the wall network shrinks away from the $\qsubrm{N}{dw}\propto t^{-1}$ law. To quantify this we   obtain an estimate for the scaling exponent, $\gamma$, defined from the average via $\overline{{N}}\qsubrm{}{dw}\propto t^{-\gamma}$, where the average is taken over all realizations in the ensemble. In \fref{fig:gammaevol} we plot the evolution of the scaling exponents (they are computed in bins of width $\qsubrm{\delta t}{bin} = 10$, beginning at $\log_{10}t=1.8$).  Again, it is clear that increasing the internal density $\rho_0$ has a substantial effect on the value of the scaling exponent. We should note that this ``constant'' scaling exponent parameterization is not totally un-ambiguous. For example, the lines in \fref{fig:gammaevol}(b) with large $\rho_0$ are not particularly straight. That said, constant-$\gamma$ is a reasonable indicator of modified scaling dynamics.

 It should be noted that forever increasing $\rho_0$ will not forever decrease $\gamma$: our results seem to suggest the existence of some critical exponent, $\qsubrm{\gamma}{crit}$ say, whose value depends on the box size. We find $\qsubrm{\gamma}{crit} \sim 0.3$ for the $L=256$ systems, $\qsubrm{\gamma}{crit} \sim 0.5$ for the $L=512$ systems, and  $\qsubrm{\gamma}{crit} \sim 0.8$ for the $L=1024$ systems. The value of the critical exponent is  dependant upon the relative sizes of the distribution of impurities and the box size, and  is interpretable as the manifestation of a scale other than the box-size being put into the system. One should expect that as the size of the box is taken to infinity, the $\propto t^{-1}$ scaling law is restored: however, once there are structures or impurities which take up an appreciable fraction of the total volume available, the scaling law becomes substantially modified.

%{\renewcommand{\arraystretch}{2}
%\begin{table}[!t]
%\begin{center}
%\begin{tabular}{|c||c|c|c|c|c|}  \hline 
%$\rho_0$& 0&1 &5 &15 & 20 \\ \hline
%$\gamma_{(L=256)}$ & 1.02& 0.56& 0.38& 0.29 & 0.29\\ \hline
%$\gamma_{(L=512)}$ & 1.05& 0.80& 0.64& 0.56 & 0.57\\ \hline
%$\gamma_{(L=1024)}$ & 1.03& 0.92& 0.84& 0.79 & 0.80 \\ \hline
%\end{tabular}
%\end{center}
%\caption{Scaling exponents $\gamma$, defined via $\overline{{N}}\qsubrm{}{dw}\propto t^{-\gamma} $, as computed from the average of 20 realizations of the  evolution of the number of domain walls from random initial conditions, for a range of values of the internal density $\rho_0$ of the impurities (the properties of which are given in the caption of \fref{fig:nwalls}). The   rows correspond to the different physical lengths of the boxes (shown as the subscript). }
%\label{tab_scaling-exponents}
%\end{table}%
%}

\subsection{Scalar radiation inside impurities}
In addition to understanding the evolution of the scalar field outside the impurities, the  next question we ask is: ``what happens inside the impurity?''. Even though we are anchoring the value of the impurity density, this does not mean we are anchoring the value of the scalar field inside the impurity. What this means is that the scalar field could   evolve inside the impurity. It is furthermore conceivable that as domain walls outside the impurity annihilate, energy could be injected into the impurity. We can address these ``conjectures'' with numerical experiments.

\begin{figure*}[!t]
      \begin{center}
      	 {\subfigure[\,$t = 12.8$ ]{\includegraphics[scale=0.2]{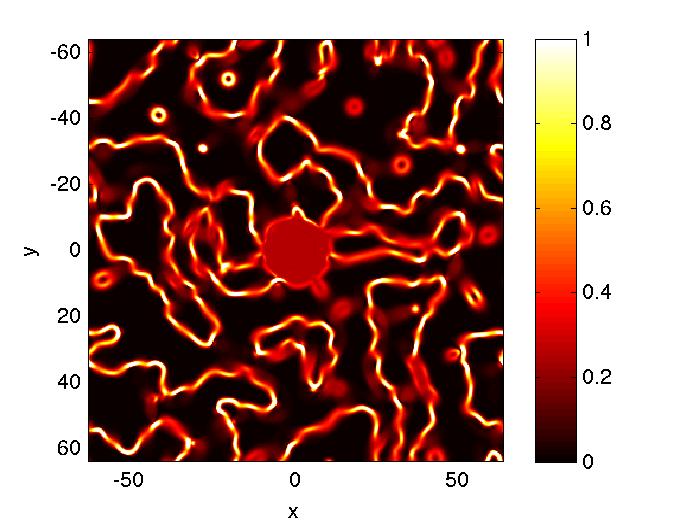}}}
       	 {\subfigure[\, $t=25.6$]{\includegraphics[scale=0.2]{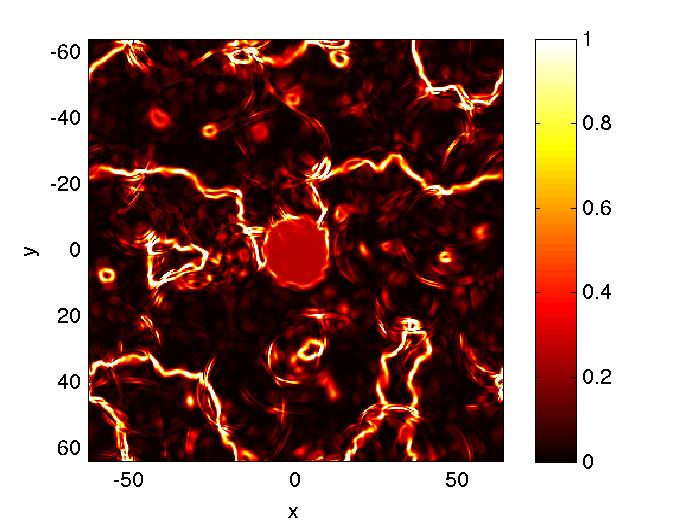}}}
      	 {\subfigure[\, $t=51.2$]{\includegraphics[scale=0.2]{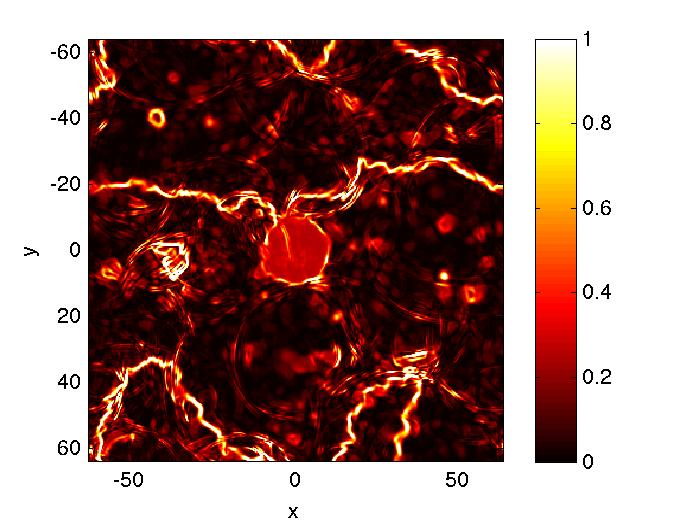}}}
      \end{center}
\caption{ Images of the energy density at various times (the times are logarithmically spaced) for a system evolving from random initial conditions with a single impurity of radius $r_0 = 10$ at the origin, with internal density $\rho_0 = 5$. This simulation has relatively small physical length $L = 128$, but this is enough to illustrate our point.} \label{fig:neck_rand-enin}
\end{figure*}

\begin{figure*}[!t]
      \begin{center}
       	 {\subfigure[\, $t = 12.8$]{\includegraphics[scale=0.2]{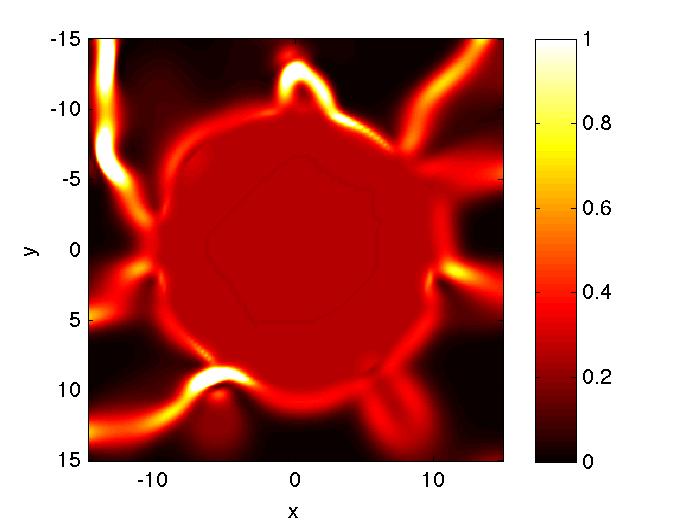}}}
       	 {\subfigure[\, $t=25.6$]{\includegraphics[scale=0.2]{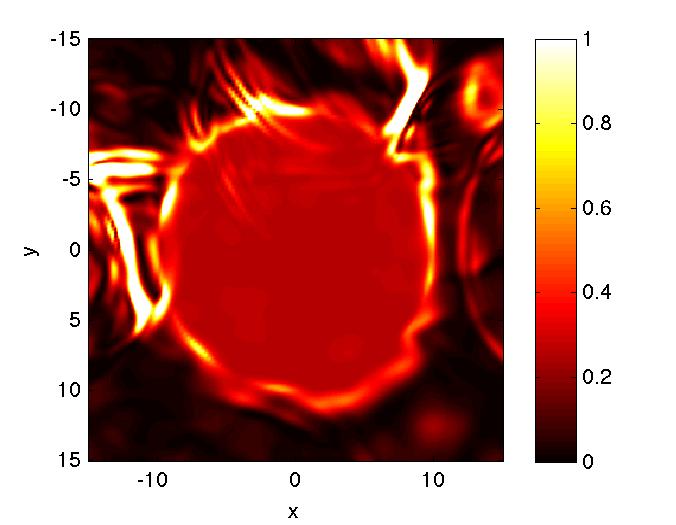}}}
      	 {\subfigure[\,$t=51.2$ ]{\includegraphics[scale=0.2]{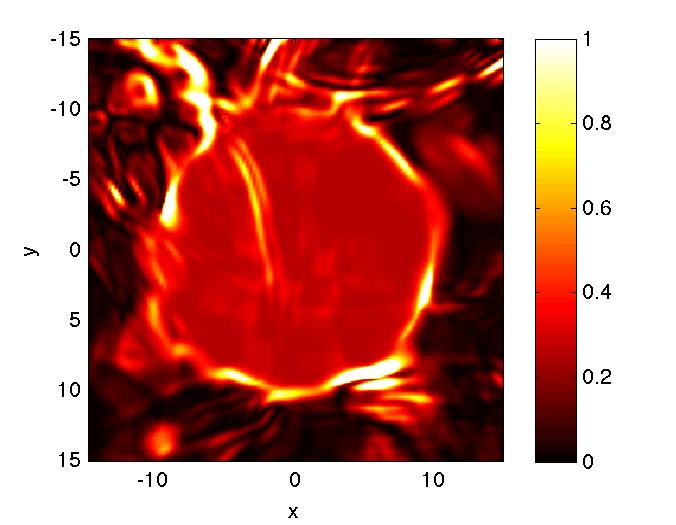}}}
      \end{center}
\caption{  Images of precisely the same simulation we presented in \fref{fig:neck_rand-enin}, but deliberately zoomed in to focus on the impurity. The important thing we want to draw attention to is  the changing energy density \textit{inside} the impurity. } \label{fig:neck_rand-enin-zoom}
\end{figure*}

\begin{figure}[!t]
      \begin{center}
{\includegraphics[scale=0.7]{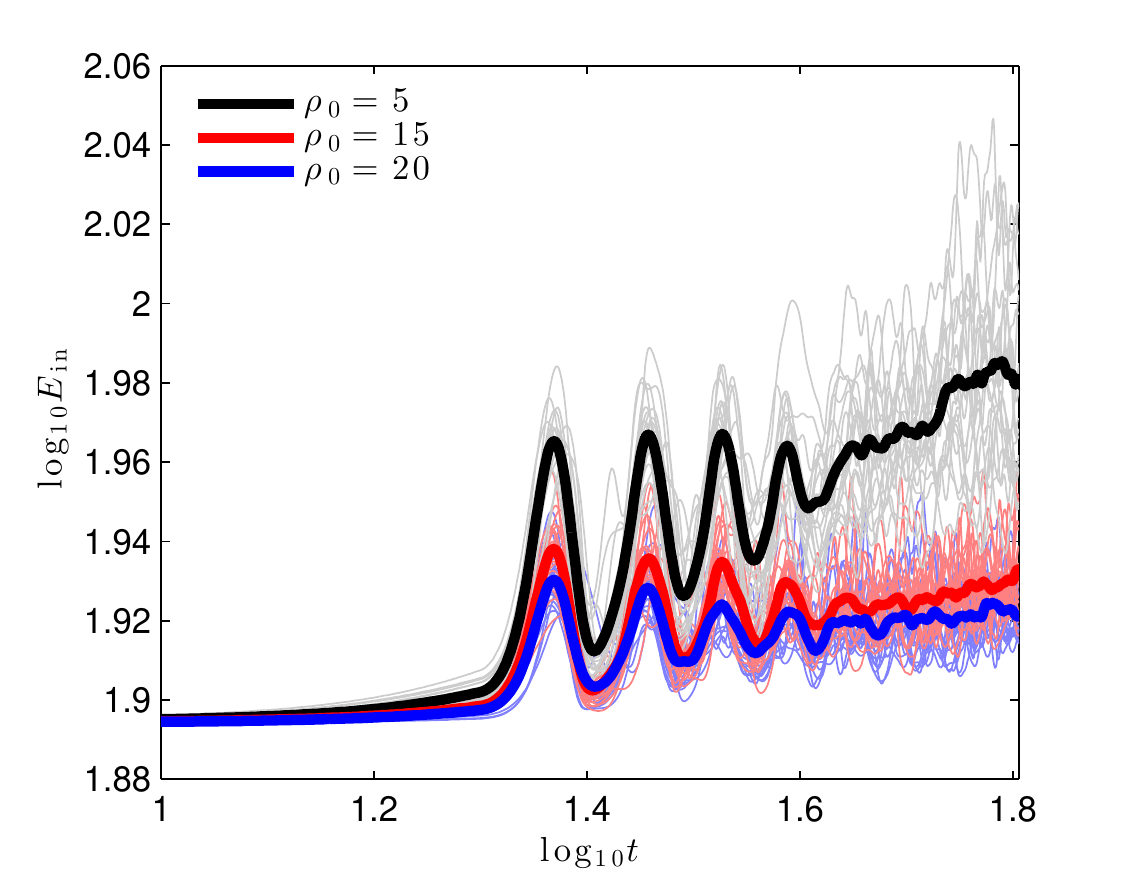}}
      \end{center}
\caption{Evolution of the energy inside impurities of size $r_0=10$ and with various internal densities as shown in the legend; there is only one impurity at the origin. The thin lines denote the evolution of each of the members of the 20 realizations, and the thick lines are the averages. } \label{fig:neck_rand-enplot}
\end{figure}

We setup a single impurity of radius $r_0 = 10$ at the origin, use the same random initial conditons described in the previous section, and evaluate the evolution of energy inside the impurity, $\qsubrm{E}{in}$. We define ``inside the impurity'' to be the regions of space where the impurity density is at least half of the internal density $\rho_0$. In \fref{fig:neck_rand-enin} we give images of the evolution of the energy density for a single impurity at the centre of the simulation (the exact properties are given in the caption of the figure). It is clear from these images that the domain walls have been evolving, decaying, and are pinned to the impurity. Moreover, in \fref{fig:neck_rand-enin-zoom} we present a zoom in of the energy density, centering on the region where the impurity resides. It is clear that the energy density within the impurity is in turmoil. There appears to be scalar radiation propagating an energy density comparable in magnitude to that on the domain wall itself. In \fref{fig:neck_rand-enplot} we plot the evolution of the energy inside the impurity for a range of internal densities (all simulations are run with 20 realizations, on boxes of physical length $L = 128$). Increasing the internal density has the effect of decreasing the scalar radiation inside the impurity -- this is relatively obvious since it will be harder for the scalar field to move in an extremely heavy environment.

\subsection{Necklace formation}
We are also interested in finding domain wall necklaces that form from random initial conditions. To find these we need to run the randomly initalized simulations   past the light crossing time $\qsubrm{t}{lx}$; we run to $t = 4\qsubrm{t}{lx}$, and use lattices  with physical length $L=64$.  

\begin{figure}[!t]
      \begin{center}
      	 {\subfigure[\, No necklace has formed]{\includegraphics[scale=0.3]{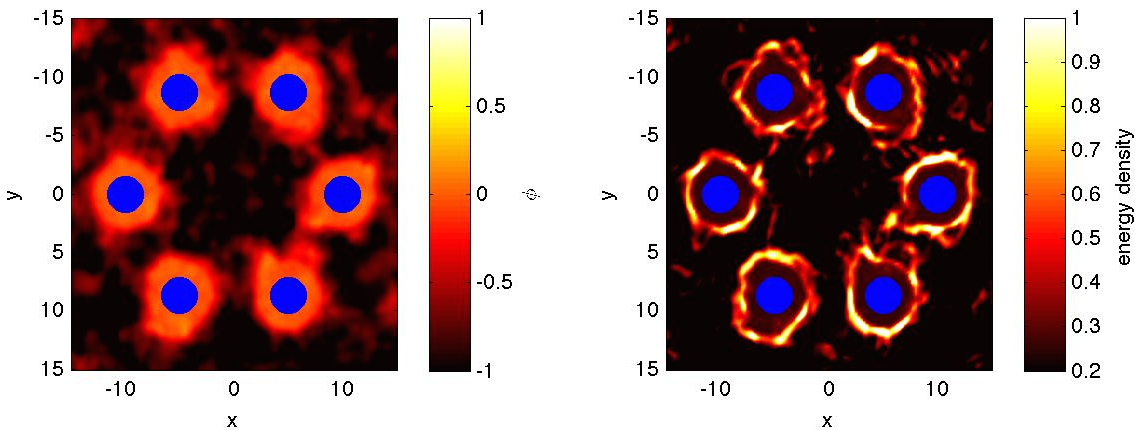}}}
      	 {\subfigure[\,  Necklace has formed ]{\includegraphics[scale=0.3]{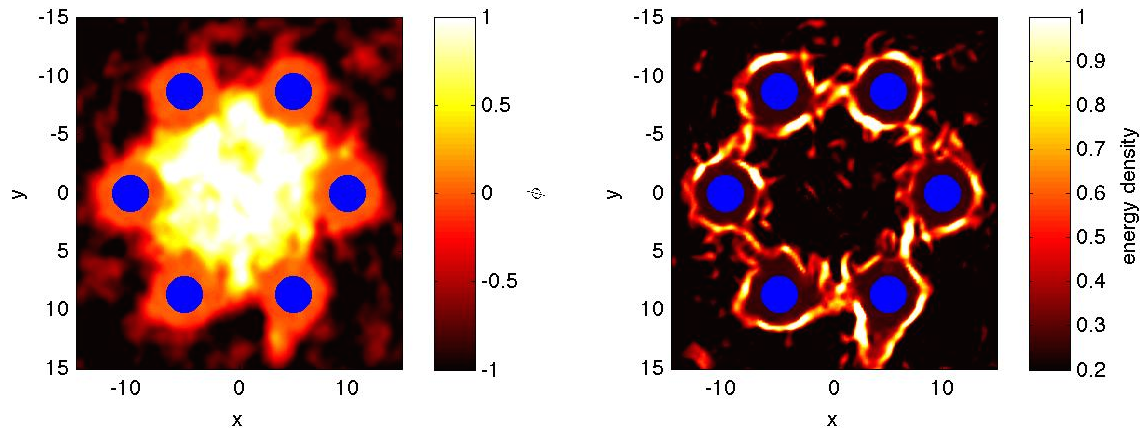}}}
      \end{center}
\caption{ Figures from two members of the realizations at $t = 4\qsubrm{t}{lx}$, from the evolution of domain walls with random initial conditions. On the left panels we give the scalar field, and on the right the energy density. Each of the $N=6$ impurities   have radius $r_0=3$, and sit on a circle of radius $R_0=10$. We have deliberately zoomed in to focus   on the region containing the impurities which are marked on with blue circles. On the top row we present a system which has not formed a domain wall necklace, and on the bottom row we an example of a system which has formed a domain wall necklace from random initial conditions. } \label{fig:neck_rand}
\end{figure}

In \fref{fig:neck_rand} we present the field configurations at $t = 4\qsubrm{t}{lx}$ for two examples of members of the ensemble of realizations. In the top line it is clear that a domain wall necklace has not formed, but in the bottom line one can clearly observe the walls linking the impurities forming a necklace. These had $N=6$  impurities, each of which have the same size properties as those in Section \ref{sec:dwneck}. When we take $N=3$ we do not find any necklace configurations, in accord with the results in Section \ref{sec:dwneck}.

\begin{figure*}[!t]
      \begin{center}
         	{\includegraphics[scale=0.3]{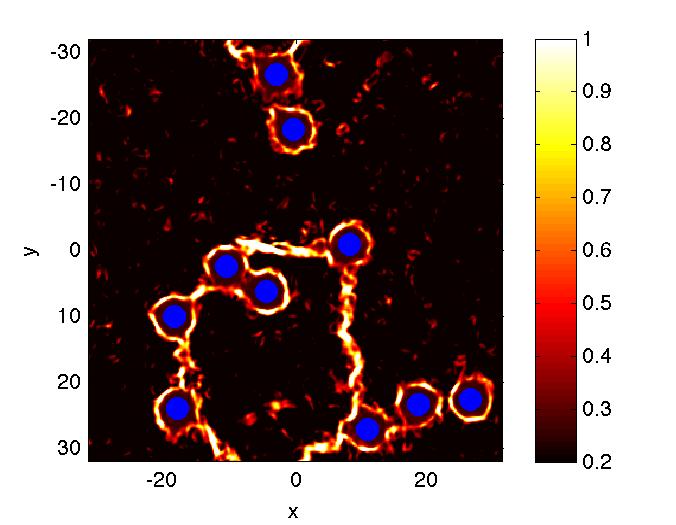}}
      	{\includegraphics[scale=0.3]{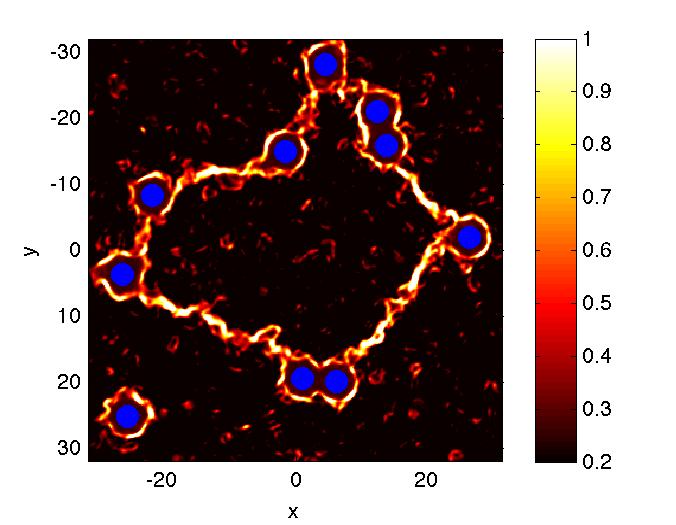}}
      \end{center}
\caption{ Examples of necklaces formed from random initial conditions, for randomly located impurities. We have marked   the locations of the impurities with blue circles onto these plots of the energy density. We have also applied a minimum ``threshold'' for the colour scheme: everywhere the energy density is lower than 0.2, the colour is set to ``black''.} \label{fig:neck_rand-rand}
\end{figure*}

Rather than put the impurities onto a circle, which is what we have done in the majority of the paper, it is also interesting to study what happens in ``random'' scenarios, by which we mean that the  impurities are placed in random locations on the simulation grid. We use $r_0=3$ to specify the size of the impurities, and run to $t = 6\qsubrm{t}{lx}$. In \fref{fig:neck_rand-rand} we give   images from two systems   which we found   formed a necklace in a very asymmetric scenario.

\section{Discussion}
\label{sec:discussion}
In this paper we have begun a systematic study of various aspects of relativistic domain walls in the presence of static symmetry-restoring impurities. The presence of the impurities gives rise to a variety of interesting phenomena, which we summarise below. Our main results are
\begin{itemize}
\item We have shown that domain walls pin to impurities.
\item Domain wall necklaces can be energetically preferred configurations.  
\item The presence of impurities   modifies the ``usual'' $\qsubrm{N}{dw} \propto t^{-1}$ scaling law.
\end{itemize}

We have also provided evidence for necklace formation from random initial conditions. To be precise, we have shown that systems evolve in such a way that domain walls link the set of impurities: this is true when the impurities are spaced regularly on a circle, and when they are randomly distributed.

There are a few limitations to the present work, the most noteworthy of which is that we have not given the impurities any dynamics; this can be justified for preliminary studies such as those outlined here.  In the field theory context, this is justified by supposing that the impurities are extremely massive fields, compared to the mass of the domain walls. In a cosmological context, this is is satisfactory as a first approximation: there the impurities are supposed to be structure  in the Universe (galaxies, clusters, voids, etc) and they have slow dynamics on a relativistic time-scale. Another limitation  which we imposed  for computational brevity, is that we only worked in 2 spatial dimensions. Increasing the numerics to 3D is simple, but computationally very expensive. We expect many of the ideas to carry over; rather than talk of ``domain wall necklaces'', we expect to find ``domain wall nets''.

Suppose that one ``lived'' on the surface of an impurity. The pinned domain wall would appear to such an observer as a point of energy. If the impurity were a 3D sphere rather than a 2D circle, then the pinned wall would appear as a (closed) line of energy on the impurities surface.

We hope that this paper will encourage further studies of domain walls in the presence of impurities. We have studied only the simplest possible domain wall forming theory, and there are many more on the market -- notably those with junctions and conserved charges mentioned in the Introduction.

%% file: symm_appendix.tex
\section{Numerical implementation}
\label{sec:append-numerics}
For completeness we shall describe in detail how to numerically solve the field equations.

Space and time are both discretized onto a lattice, with grid-spacings between the sites $\Delta x$ and $\Delta t$ respectively. We use $\phi^T_{i,j}$ to denote the value of a scalar $\phi$ at time-step number $T$, and at the spatial location whose coordinates on the lattice are $(i, j)$. Derivatives are discretized to second order in time and space using finite differences (as opposed to other schemes, like Fast Fourier Transforms); higher order finite difference schemes are available, but we stick with second order since it is adequate for our purposes (it is also the   computationally cheapest scheme). 

The finite difference scheme is implemented via the following discretization of first derivatives evaluated at the lattice site whose coordinates are $(i,j)$ and at time-step $T$:
\bse
\bea
\label{eq:first-t-disc}
\pd{\phi^T_{i,j}}{t} = \frac{\phi^{T+1}_{i,j} - \phi^{T-1}_{i,j}}{2\Delta t},  
\eea
\bea
\pd{\phi^T_{i,j}}{x} = \frac{\phi^T_{i+1,j} - \phi^T_{i-1,j}}{2\Delta x}.
\eea
\ese
Second derivatives are discretized as
\bse
\bea
\label{eq:sec-t-disc}
\pd{^2\phi^T_{i,j}}{t^2} = \frac{\phi^{T+1}_{i,j} - 2 \phi^T_{i,j} + \phi^{T-1}_{i,j}}{\left(\Delta t\right)^2},
\eea
\bea
\pd{^2\phi^T_{i,j}}{x^2} = \frac{\phi^T_{i+1,j} - 2 \phi^T_{i,j} + \phi^T_{i-1,j}}{\left(\Delta x\right)^2}.
\eea
\ese
Similar expressions are obtained for the derivatives in the $y$-direction, so that in particular the Laplacian in 2D is discretized as
\bea
\nabla^2\phi^T_{i,j} = \frac{\phi^T_{i+1,j} + \phi^T_{i-1,j} +\phi^T_{i,j+1} + \phi^T_{i,j-1} - 4 \phi^T_{i,j}}{\left( \Delta x\right)^2}.
\eea
Care needs to be taken at the boundaries.  For a box where the lattice coordinates take on discrete values $0 \leq i < \qsubrm{i}{max}$, periodic boundary conditions are implemented via the identifications
\bea
\phi^T_{-1,j} = \phi^T_{\qsubrm{i}{max}-1,j},\qquad \phi^T_{\qsubrm{i}{max},j} = \phi^T_{0,j},
\eea
with equivalent identifications in the $j$-direction.

To obtain an  algorithm to update the value of the scalar field as time increments, it is useful to write the equation of motion (\ref{eom-damoubg}) as
\bea
\label{eq:append_eom}
\ddot{\phi}^T_{i,j} + \alpha\dot{\phi}^T_{i,j} = \mathcal{E}_{i,j}^T,
\eea
in which
\bea
\mathcal{E}^T_{i,j} \defn \nabla^2\phi^T_{i,j} - V_{,\phi^T_{i,j}}
\eea
can be evaluated at a given lattice site. Using (\ref{eq:first-t-disc}) and (\ref{eq:sec-t-disc}) for the first and second time derivatives respectively, the equation of motion (\ref{eq:append_eom}) rearranges to give
\bea
\phi^{T+1}_{i,j} = \frac{1}{1+ \tfrac{1}{2}{\alpha}\Delta t}\left[\left( \Delta t\right)^2\mathcal{E}_{i,j}^T - \left(  1-\tfrac{1}{2} \alpha\Delta t\right)\phi^{T-1}_{i,j} + 2 \phi_{i,j}^T\right].
\eea
This provides a scheme for using the current and previous values of the scalar to obtain the next value. This is known as a leapfrog scheme. On a practical note, this means that only two time-steps worth of data   needs to be stored in memory: the ``new'' value $\phi^{T+1}_{i,j}$ overwrites the ``previous'' value $\phi^{T-1}_{i,j}$.

%% file: symm_paper.bbl
\providecommand{\href}[2]{#2}\begingroup\raggedright\endgroup